\documentclass[twocolumn]{aastex62}
\newcommand{\gt}{>}
\usepackage{amsmath}

\accepted{ApJ}

\shorttitle{STUDIES IV: Spatial Clustering and Halo Masses of 450-\lowercase{$\micron$} Sources}
\shortauthors{C.-F. Lim et al.}

\let\oldAA\AA
\renewcommand{\AA}{\text{\normalfont\oldAA}}

\graphicspath {{figures/}}
\begin{document}

\title{SCUBA-2 ULTRA DEEP IMAGING EAO SURVEY (STUDIES) IV: Spatial clustering and halo masses of 450-$\micron$-selected sub-millimeter galaxies}

\correspondingauthor{Chen-Fatt Lim}
\email{chenfatt.lim@gmail.com}

\author{Chen-Fatt Lim}
\affiliation{Academia Sinica Institute of Astronomy and Astrophysics (ASIAA), No. 1, Section 4, Roosevelt Road, Taipei 10617, Taiwan}
\affiliation{Graduate Institute of Astrophysics, National Taiwan University, Taipei 10617, Taiwan}

\author{Chian-Chou Chen}
\affiliation{Academia Sinica Institute of Astronomy and Astrophysics (ASIAA), No. 1, Section 4, Roosevelt Road, Taipei 10617, Taiwan}
\affiliation{European Southern Observatory, Karl Schwarzschild Strasse 2, Garching, Germany}

\author{Ian Smail}
\affiliation{Centre for Extragalactic Astronomy, Department of Physics, Durham University, South Road, Durham, DH1 3LE, UK}

\author{Wei-Hao Wang}
\affiliation{Academia Sinica Institute of Astronomy and Astrophysics (ASIAA), No. 1, Section 4, Roosevelt Road, Taipei 10617, Taiwan}

\author{Wei-Leong Tee}
\affiliation{Department of Physics, University of Arizona, Tucson, Arizona 85721, USA}
\affiliation{Academia Sinica Institute of Astronomy and Astrophysics (ASIAA), No. 1, Section 4, Roosevelt Road, Taipei 10617, Taiwan}

\author{Yen-Ting Lin}
\affiliation{Academia Sinica Institute of Astronomy and Astrophysics (ASIAA), No. 1, Section 4, Roosevelt Road, Taipei 10617, Taiwan}

\author{Douglas Scott}
\affiliation{Department of Physics \& Astronomy, University of British Columbia, BC, Canada}

\author{Yoshiki Toba}
\affiliation{Department of Astronomy, Kyoto University, Kitashirakawa-Oiwake-cho, Sakyo-ku, Kyoto 606-8502, Japan}
\affiliation{Academia Sinica Institute of Astronomy and Astrophysics (ASIAA), No. 1, Section 4, Roosevelt Road, Taipei 10617, Taiwan}
\affiliation{Research Center for Space and Cosmic Evolution, Ehime University, 2-5 Bunkyo-cho, Matsuyama, Ehime 790-8577, Japan}

\author{Yu-Yen Chang}
\affiliation{Academia Sinica Institute of Astronomy and Astrophysics (ASIAA), No. 1, Section 4, Roosevelt Road, Taipei 10617, Taiwan}

\author{YiPing Ao}
\affiliation{Purple Mountain Observatory and Key Laboratory for Radio Astronomy, Chinese Academy of Sciences, Nanjing 210033, People's Republic of China}

\author{Arif Babul}
\affiliation{Department of Physics and Astronomy, University of Victoria, Elliott Building, 3800 Finnerty Road, Victoria, BC V8P 5C2, Canada}

\author{Andy Bunker}
\affiliation{Department of Physics, University of Oxford, Denys Wilkinson Building, Keble Road, Oxford OX13RH, UK}
\affiliation{Visitor, Kavli Institute for the Physics and Mathematics of the Universe (WPI), the University of Tokyo Institutes for Advanced Study, Japan}

\author{Scott C. Chapman}
\affiliation{Department of Physics and Astronomy, University of British Columbia, 6225 Agricultural Road, Vancouver, BC, V6T 1Z1, Canada}
\affiliation{National Research Council, Herzberg Astronomy and Astrophysics, 5071 West Saanich Road, Victoria, BC, V9E 2E7, Canada}
\affiliation{Department of Physics and Atmospheric Science, Dalhousie University, Halifax, NS B3H 4R2, Canada}

\author{David L Clements}
\affiliation{Department of Physics, Blackett Lab, Imperial College, Prince Consort Road, London, SW7 2AZ, UK}

\author{Christopher J. Conselice}
\affiliation{University of Nottingham, School of Physics \& Astronomy, Nottingham, NG7 2RD, UK}

\author{Yu Gao}
\affiliation{Purple Mountain Observatory and Key Laboratory for Radio Astronomy, Chinese Academy of Sciences, Nanjing 210033, People's Republic of China}
\affiliation{Department of Astronomy, Xiamen University, Xiamen, Fujian 361005, People's Republic of China}

\author{Thomas R. Greve}
\affiliation{Department of Physics and Astronomy, University College London, Gower Street, London WC1E 6BT, UK}
\affiliation{Cosmic Dawn Center}

\author{Luis C. Ho}
\affiliation{Kavli Institute for Astronomy and Astrophysics, Peking University, Beijing 100871, People’s Republic of China}
\affiliation{Department of Astronomy, School of Physics, Peking University, Beijing 100871, People’s Republic of China}

\author{Sungwook E. Hong}
\affiliation{Natural Science Research Institute, University of Seoul, 163 Seoulsiripdaero, Dongdaemun-gu, Seoul 02504, Republic of Korea}

\author{Ho Seong Hwang}
\affiliation{Korea Astronomy and Space Science Institute, 776 Daedeokdae-ro, Yuseong-gu, Daejeon 34055, Republic of Korea}

\author{Maciej Koprowski}
\affiliation{Institute of Astronomy, Faculty of Physics, Astronomy and Informatics, Nicolaus Copernicus University, Grudziadzka 5, 87-100 Torun, Poland}

\author{Micha{\l} J.~Micha{\l}owski}
\affiliation{Astronomical Observatory Institute, Faculty of Physics, Adam Mickiewicz University, 60-286 Pozna$\acute{n}$, Poland}

\author{Hyunjin Shim}
\affiliation{Department of Earth Science Education, Kyungpook National University, Deagu 41566, Republic of Korea}

\author{Xinwen Shu}
\affiliation{Department of Physics, Anhui Normal University, Wuhu, Anhui, 241000, People’s Republic of China}

\author{James M. Simpson}
\affiliation{Academia Sinica Institute of Astronomy and Astrophysics (ASIAA), No. 1, Section 4, Roosevelt Road, Taipei 10617, Taiwan}

\begin{abstract} 
We analyze an extremely deep 450-$\micron$ image ($1\sigma=0.56$\,mJy\,beam$^{-1}$) of a $\simeq 300$\,arcmin$^{2}$ area in the CANDELS/COSMOS field as part of the SCUBA-2 Ultra Deep Imaging EAO Survey (STUDIES). We select a robust (signal-to-noise ratio $\geqslant 4$) and flux-limited ($\geqslant 4$\,mJy) sample of 164 sub-millimeter galaxies (SMGs) at 450 $\micron$ that have $K$-band counterparts in the COSMOS2015 catalog identified from radio or mid-infrared imaging. Utilizing this SMG sample and the 4705 $K$-band-selected non-SMGs that reside within the noise level $\leqslant 1$\,mJy\,beam$^{-1}$ region of the 450-$\micron$ image as a training set, we develop a machine-learning classifier using $K$-band magnitude and color-color pairs based on the thirteen-band photometry available in this field. We apply the trained machine-learning classifier to the wider COSMOS field (1.6\,deg$^{2}$) using the same COSMOS2015 catalog and identify a sample of 6182 SMG candidates with similar colors. The number density, radio and/or mid-infrared detection rates, redshift and stellar mass distributions, and the stacked 450-$\micron$ fluxes of these SMG candidates, from the S2COSMOS observations of the wide field, agree with the measurements made in the much smaller CANDELS field, supporting the effectiveness of the classifier. Using this 450-$\micron$ SMG candidate sample, we measure the two-point autocorrelation functions from $z=3$ down to $z=0.5$. We find that the SMG candidates reside in halos with masses of $\simeq (2.0\pm0.5) \times10^{13}\,h^{-1}\,\rm M_{\sun}$ across this redshift range. We do not find evidence of downsizing that has been suggested by other recent observational studies.

\end{abstract}


\section{Introduction} \label{sec:Introduction}

Over the past two decades, a class of far-infrared luminous galaxies has been discovered at sub-millimeter wavelengths. The extreme infrared luminosities observed in these sub-millimeter galaxies (SMGs) suggest that they are dusty and considered to be among the most intensively star-forming sources in the Universe \citep{Smail:1997aa, Barger:1998aa, Barger:1999aa, Hughes:1998aa, Eales:1999aa}. SMGs appear to have a redshift distribution peaking at $z \simeq 2.5$ with the majority of them at $z=1.5$--3.5 \citep{Barger:2000aa, Chapman:2003ab, Chapman:2005aa, Pope:2006aa, Aretxaga:2007aa, Wardlow:2011aa, Michaowski:2012ab, Yun:2012aa, Simpson:2014aa, Simpson:2017ab, Chen:2016ab, Koprowski:2016aa, Danielson:2017aa, Dunlop:2017aa, Michaowski:2017aa, Dudzeviciute:2020aa, Stach:2019aa}, occupying the same putative peak epoch of unobscured star formation \citep{Madau:2014aa} and active galactic nucleus (AGN) activity \citep{Schmidt:1995aa, Hasinger:2005aa, Wall:2008aa, Assef:2011aa}. The total infrared luminosities ($L_{\rm IR}$; 8--1000\,$\micron$) of SMGs are similar to local ultra-luminous infrared galaxies (ULIRGs; \citealt{Sanders:1988aa, Sanders:1996aa, Farrah:2001aa, Armus:2009aa}), reaching values greater than a few times $10^{12}\,\rm L_{\sun}$ or even higher than $10^{13}\,\rm L_{\sun}$ for some of the brightest sources. This corresponds to star-formation rates (SFRs) ranging from around $100\,\rm M_{\sun}\,yr^{-1}$ to more than $1000\,\rm M_{\sun}\,yr^{-1}$ \citep{Hainline:2011aa, Barger:2012aa, Swinbank:2014aa, da-Cunha:2015aa, Simpson:2015aa, Dudzeviciute:2020aa}. 

Violent gas accretion, potentially induced by mergers, is the most likely explanation to date for the intensive star formation of SMGs \citep{Frayer:1998aa, Conselice:2003aa, Greve:2005aa, Tacconi:2008aa, Engel:2010aa, Swinbank:2010aa, Alaghband-Zadeh:2012aa, Menendez-Delmestre:2013aa, Chen:2015aa, Koprowski:2016aa, Michaowski:2017aa, Chang:2018aa}. Large amounts of gas accretion can result in a short-lived starburst possibly followed by a quasar phase. Feedback mechanisms from star formation or black-hole accretion could have injected sufficient energy to heat the remaining gas, or expel it from the galaxy to prevent further star formation \citep{Silk:1998aa, Fabian:1999aa, Trayford:2016aa}. This scenario may be responsible for the formation of the most massive ($M_{\ast}>$10$^{11}$\,$\rm M_{\sun}$) elliptical galaxies in the local Universe \citep{Lilly:1999aa, Hopkins:2005aa, Simpson:2014aa, Toft:2014aa, Dudzeviciute:2020aa, Rennehan:2020aa}. Therefore, the cosmological evolution of SMGs is crucial for our understanding of the formation of massive galaxies in the Universe. 

Comparison of clustering measurements with dark matter simulations can provide constraints on the masses of dark matter halos that a given galaxy population resides in \citep{Peebles:1980aa} and further trace the evolution of the given galaxy population. Previous clustering analyses of SMGs identified in shorter (250--500\,$\micron$, \citealt{Cooray:2010aa, Maddox:2010aa, Mitchell:2012aa, vanKampen:2012aa, Amvrosiadis:2019aa}) and longer (850--1100\,$\micron$, \citealt{Scott:2002aa, Webb:2003ab, Weis:2009aa, Lindner:2011aa, Williams:2011aa, Hickox:2012aa, Wilkinson:2017aa, An:2019aa}) sub-millimeter wavebands have revealed that SMGs reside in high-mass (10$^{12}$--10$^{13}$\,$h^{-1}$\,$\rm M_{\sun}$) dark matter halos. These values are also consistent with previous estimates from a sample of obscured starburst galaxies reported by \cite{Bethermin:2014aa}, in which they used a combined $BzK$ color criterion and $\emph{Herschel}$/PACS data to study the clustering signal of obscured starburst galaxies at $1.5<z<2.5$ as a function of their physical parameters. These results suggest that SMGs may be the progenitors of massive elliptical galaxies in the local Universe \citep{Hughes:1998aa, Eales:1999aa, Swinbank:2006aa, Targett:2011aa, Miller:2018aa, Wang:2019aa}. However, many of these previous studies were limited by either the modest samples of SMGs ($\simeq100$ sources), or a lack of reliable identifications and redshift measurements which makes their estimated clustering signals highly uncertain. 

More precise determinations of the clustering properties with sizable SMG samples have been performed by \cite{Chen:2016ab}, \cite{Wilkinson:2017aa}, \cite{Amvrosiadis:2019aa}, and \cite{An:2019aa}. \cite{Chen:2016ab} identified a sample of $\simeq 3000$ faint SMGs ($S_\mathrm{850\,\micron}$ $<$ 2\,mJy) using a color selection technique and compared their clustering properties with other galaxy populations in the redshift range $1<z<5$. \cite{Wilkinson:2017aa} performed a clustering analysis using a sample of $\simeq 600$ 850-$\micron$-selected SMGs in the UKIDSS Ultra Deep Survey (UDS) field in the redshift interval $1<z<3$. \cite{Amvrosiadis:2019aa} studied the clustering properties for a sample of $\simeq 120,000$ $\emph{Herschel}$-selected SMGs with flux densities $S_\mathrm{250\,\micron}>30$\,mJy in low ($z<0.3$) and high ($1<z<5$) redshift intervals. \cite{An:2019aa} identified $\simeq 7000$ potential 850-$\micron$ SMGs in the COSMOS field based on a radio+machine-learning method trained on the ALMA-identified sample \citep{An:2018aa} and studied their clustering properties. 

These aforementioned recent studies were able to measure the clustering signals from SMGs as a function of redshift. \cite{Wilkinson:2017aa} found that the clustering appears to exhibit tentative evolution with redshift, such that the SMG activity seems to be shifting to less massive halos at lower redshift $z=1$--2 and consistent with the downsizing scenario \citep{Cowie:1996aa, Magliocchetti:2014aa, Rennehan:2020aa} that the contribution of luminous sources dominates in the early Universe, whereas the growth of the less luminous ones continues at lower redshifts. In contrast, \cite{Chen:2016ab}, \cite{Amvrosiadis:2019aa}, and \cite{An:2019aa} did not find such a trend, suggesting that SMGs reside in a typical halo mass of about $10^{13}\,h^{-1}\,\rm M_{\sun}$ across the redshift range $1<z<5$. The discrepancies in the lower redshift bins could be simply caused by the measurement uncertainties (uncertain identifications and/or poor redshift measurements), or by the different methodologies that are adopted in the clustering analyses, where \cite{Wilkinson:2017aa} relied on the cross-correlation technique with an abundant $K$-band selected sample, while \cite{Chen:2016ab}, \cite{Amvrosiadis:2019aa}, and \cite{An:2019aa} adopted an auto-correlation technique. 

However, these studies did not probe the clustering signals in a key redshift range ($0.3<z<1.0$; cosmic time ranges from 6.0--10.4\,Gyr) in which the downsizing effect, if exists, is expected to increase. This is likely caused by the longer-wavelength observations being more sensitive to high-redshift sources. The traditional 850-$\micron$ selection allows us to measure clustering at $z>1$ \citep{Chen:2016ab, Wilkinson:2017aa, An:2019aa}, whereas the studies based on $\emph{Herschel}$ at shorter wavelengths (e.g., 250-$\micron$; \citealt{vanKampen:2012aa, Amvrosiadis:2019aa}) are mainly sensitive to brightest low-redshift sources ($S_\mathrm{250\,\micron}>30$\,mJy) due to their positive K-correction. Using spectral energy distribution (SED) template fitting on the far-infrared photometry to estimate redshifts for sources without optical counterparts, \cite{Amvrosiadis:2019aa} extended the $\emph{Herschel}$-based clustering studies to $z>1$, finding results consistent with those obtained from the 850-$\micron$ selection. However, the large redshift uncertainties ($\simeq 0.3$ for $z=1$; \citealt{Amvrosiadis:2019aa}) means that they cannot meaningfully measure the clustering signals in the low-redshift regime.

Observations at mid-infrared (e.g., 24\,$\micron$) can be used to overcome the aforementioned selection bias and probe the clustering signals in the key redshift range of $0.3<z<1.0$ \citep{Gilli:2007aa, Magliocchetti:2008aa, Starikova:2012, Dolley:2014aa, Solarz:2015aa}. These studies found relatively lower clustering strengths at $z<1$ with clustering length $r_{\rm 0} = 3$--$6\,h^{-1}$\,Mpc compared to the high-redshift measurements. This finding is similar to the work of \cite{Magliocchetti:2013aa} based on $\emph{Herschel}$/PACS 100-$\micron$-selected sources, in which they found clear evidence for downsizing effect at redshifts limited to $z \lesssim 1$. However, the clustering lengths also correlate with the infrared luminosities, where the galaxies with higher $L_{\rm IR}$ (higher SFRs) tend to have stronger clustering signals \citep{ Dolley:2014aa, Toba:2017aa}. The majority of the sources in the above studies at 24\,$\micron$ and 100\,$\micron$ are biased toward a fainter population with $L_{\rm IR} \simeq 10^{11}\,\rm L_{\sun}$ at $z<1$, which prevents us from making a fair comparison with the SMGs at $z>1$ that have $L_{\rm IR}>10^{12}\,\rm L_{\sun}$. 

In this paper, we base our analysis on the 450-$\micron$ data obtained from the Sub-millimeter Common User Bolometric Array-2 \citep[SCUBA-2, ][]{Holland:2013aa} camera on the 15-m James Clerk Maxwell Telescope (JCMT). The 450-$\micron$ observations allow us to obtain photometric measurements closer to the redshifted SED peak of typical SMGs ($\lambda_{\rm rest}\simeq100\,\micron$) so they provide a closer match to far-infrared luminosity selection compared to longer wavelength observations. The 450-$\micron$ observations also allow us to probe the SMGs at lower redshifts ($z\simeq1.5$), with the majority of them at $z=0.5$--3.0 \citep{Casey:2013aa, Simpson:2014aa, Bourne:2017aa, Zavala:2018aa, Lim:2020aa}. We have pushed the frontier of the 450-$\micron$ imaging by initiating a new SCUBA-2 imaging survey in the CANDELS/Cosmic Evolution Survey (COSMOS, \citealt{Scoville:2007aa}) field, called the SCUBA-2 Ultra Deep Imaging EAO Survey (STUDIES, \citealt{Wang:2017aa}). By including all the archival data, we have constructed an extremely deep single-dish image at 450\,$\micron$ ($\sigma_{450\,\micron}$ = 0.56\,mJy\,beam$^{-1}$), which is the deepest image yet obtained at 450\,$\micron$. A series of papers had been published based on STUDIES, including number counts \citep{Wang:2017aa}, stellar morphology \citep{Chang:2018aa}, and multi-wavelength properties of the sample \citep{Lim:2020aa}. In this work, we develop a machine-learning classifier based on 164 sources that have 450-$\micron$ flux density $\geqslant$ 4\,mJy, signal-to-noise ratio (S/N) $\geqslant 4$, and $K$-band counterparts from their radio or mid-infrared identifications. Our machine-learning classifier labels a sample of 6182 SMG candidates in the wider COSMOS field. We employ an auto-correlation technique on the SMG candidates to statistically estimate their clustering signal, which allows us to infer the dark-matter halo mass and to constrain the clustering evolution of SMGs from $z=3$ down to $z=0.5$.

This paper is structured as follows. In \S\ref{sec:Data}, we introduce the ancillary data in the COSMOS field, as well as the observations, data reduction techniques, source extraction procedure, and training dataset. In \S\ref{sec:Methodology}, we describe the machine-learning methodology we use for SMG candidate identification. In \S\ref{sec:Justifications}, we verify our machine-learning technique in selecting the SMG candidates. We present the comparison samples in \S\ref{sec:Comparison} and the clustering properties of SMG candidates in \S\ref{sec:Clustering}. We summarize our findings in \S\ref{sec:Summary}. Throughout this work, the standard errors of our sample distribution medians are estimated from bootstrap analysis. The term ``SMG candidates" in this work represents the machine-learned candidates of 450-$\micron$-selected SMGs, unless otherwise stated. We adopt cosmological parameters $H_{0}$ = 70 km\,s$^{-1}$\,Mpc$^{-1}$, $\Omega_{\Lambda}$ = 0.70, $\Omega_{\rm m}$ = 0.30 and $\sigma_8=0.83$ \citep{Planck:2014aa}.

\section{Data} \label{sec:Data}

	\begin{figure*}
	\includegraphics[width=0.9\paperwidth]{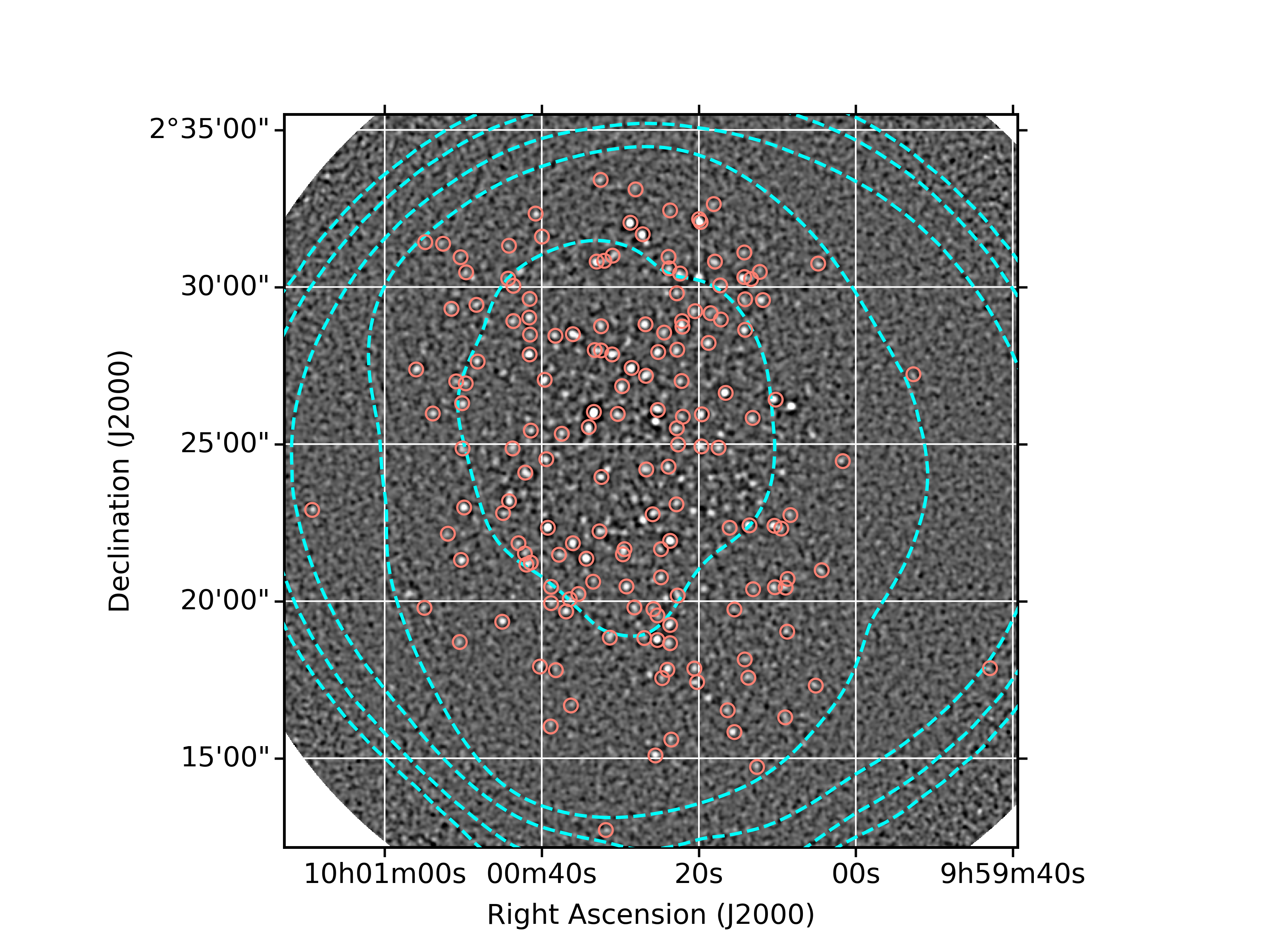}
	\caption{JCMT SCUBA-2 450-$\micron$ S/N image, showing the positions of the 164 S/N $\geqslant$ 4 and $S_\mathrm{450\,\micron}$ $\geqslant$ 4\,mJy sources with $K$-band counterparts based on the VLA and MIPS identifications (red circles). The cyan contours show the instrumental noise levels with contours at 1, 5, 9, 13, and 17\,mJy.}
	\label{fig:SnrImage}
	\end{figure*}

\subsection{Main data} 
The SCUBA-2 camera contains 5000 pixels (field of view $\simeq 45$\,arcmin$^{2}$) in each of the 450- and 850-$\micron$ detector arrays. The SCUBA-2 camera operates at 450 and 850\,$\micron$ simultaneously and provides an unprecedented mapping speed, meaning that it can efficiently survey large areas of sky at 450 and 850\,$\micron$. The beam size of SCUBA-2 is 7${\farcs}$9 at 450\,$\micron$, which is an order of magnitude smaller in area compared to far-infrared observations from \emph{Herschel} at similar wavelengths (24--$35\arcsec$). The 450-$\micron$ data presented in this paper come from three sources: STUDIES \citep{Wang:2017aa}, the data taken by \citet[][hereafter C13]{Casey:2013aa}, and the SCUBA-2 Cosmology Legacy Survey 450-$\micron$ campaign in the COSMOS field (S2CLS-COSMOS, \citealt{Geach:2013aa, Geach:2017aa}). We briefly describe these programs in turn.

STUDIES is a JCMT Large Program that aims to reach the confusion limit at 450\,$\micron$ within the CANDELS \citep{Grogin:2011aa, Koekemoer:2011aa} footprint in the COSMOS field. The mapping center of STUDIES is R.A. = 10$^{\rm h}$00$^{\rm m}$30${\rm \fs}$7 and decl. = +02$\degr$26$\arcmin$40$\arcsec$. The \texttt{CV DAISY} mapping pattern was used, resulting in an $3\arcmin$ diameter area of approximately uniform coverage of high sensitivity, with the noise level increasing outside this area as a function of radius. In this work, we only adopt the STUDIES data that was collected until April 2019, amounting to 252\,hours of exposures, about 84\% of the total allocated integration (330\,hours) for the whole project. We note that data collection for STUDIES is now 99\% complete (by April 2020) and the final version of the deeper image and source catalog will be published in the future. In addition, as part of the S2CLS project, the 450-$\micron$ S2CLS-COSMOS was observed with two \texttt{CV DAISY} maps offset by 2$\arcmin$ in declination from the central pointing of R.A. = 10$^{\rm h}$00$^{\rm m}$30${\rm \fs}$7 and decl. = +02$\degr$22$\arcmin$40$\arcsec$, with some overlap resulting in an area of $\simeq 13$\,arcmin$^{2}$ with noise level $< 5$\,mJy\,beam$^{-1}$. The total on-sky integration of S2CLS-COSMOS is 150\,hours. The survey of C13 used the \texttt{PONG}-900 scan pattern, mapping with a center of R.A. = 10$^{\rm h}$00$^{\rm m}$28${\rm \fs}$0 and decl. = +02$\degr$24$\arcmin$00$\arcsec$, resulting in a wider circular map that reaches a uniform depth over an area of approximately 700\,arcmin$^{2}$. The total on-target time of C13 is 38\,hours and this data are much shallower compared to the STUDIES and S2CLS datasets. The majority of the aforementioned observations were conducted under the best sub-millimeter weather on Mauna Kea (``Band 1,''  $\tau_{\rm 225\,GHz}$ $<$ 0.05, where $\tau_{\rm 225\,GHz}$ is the sky opacity at 225\,GHz). 

The procedure used for data reduction are similar to those in \cite{Wang:2017aa} and \cite{Lim:2020aa}. In brief, we apply the following steps. 
\begin{enumerate}
\item The raw time-stream data from SCUBA-2 were flat-fielded using the flat scans that bracket each science observation, and the data were scaled to units of pW.
\item The time streams were then assumed to be a linear combination of noise and signal from the background (atmospheric water and ambient thermal emission), as well as astronomical objects. The procedure then entered an iterative stage that attempts to fit these components with a model by using the \texttt{Dynamic Iterative Map-Maker} routine of Sub-Millimeter Common User Reduction Facility (SMURF, \citealt{Chapin:2013aa}). 
\item We then calibrated the individual reduced scans into units of flux density by using the weighted mean flux conversion factor (FCF) of ($476 \pm 95$)\,Jy\,beam$^{-1}$\,pW$^{-1}$. The adopted FCF was estimated from a subset of sub-millimeter calibrators observed under Band-1 weather during the survey campaigns and was consistent with the canonical value estimated from a wider base of calibrators \citep{Dempsey:2013aa}.
\item Individual scans were then co-added in an optimal, noise-weighted manner, by using the \texttt{MOSAIC\_JCMT\_IMAGES} recipe from the PIpeline for Combining and Analyzing Reduced Data (PICARD, \citealt{Jenness:2008aa}). 
\item To improve the detectability of faint point sources, we convolved the map with a broad Gaussian kernel of full width at half-maximum (FWHM) = $20\arcsec$, and subtracted the convolved map from the original maps to remove any large-scale structure in the sky background. We then convolved the subtracted map with a Gaussian kernel that is matched to the instrumental point-spread function (FWHM = 7${\farcs}$9, \citealt{Dempsey:2013aa}). We used the \texttt{SCUBA2\_MATCHED\_FILTER} recipe in the PICARD environment for this procedure.
\end{enumerate}

Finally, we constructed an extremely deep 450-$\micron$ image, with the STUDIES, C13, and S2CLS-COSMOS data combined (Figure \ref{fig:SnrImage}). The final image covers a sensitive region of $\simeq 300$\,arcmin$^{2}$. The instrumental noise level in the deepest regions is $\simeq 0.56$\,mJy\,beam$^{-1}$ that is roughly 16\% deeper than previous works \citep{Lim:2020aa}.

\subsection{Ancillary data} \label{subsec:AncillaryData}
We employed radio and near-/mid-infrared identifications to construct a sample of SMG counterparts used for this work (\S\ref{subsec:TrainingSample}). In the radio, we used the catalog from the Jansky Very Large Array (VLA) Large Project survey at 3\,GHz (VLA-COSMOS; \citealt{Smolcic:2017aa}). In the near-infrared, we employed the archival IRAC catalog \citep{Sanders:2007aa} obtained from the \emph{Spitzer Space Telescope}. In the mid-infrared, we generated our own 24-$\micron$ catalog using \texttt{SExtractor} \citep{Bertin:1996aa}, since the archival MIPS 24-$\micron$ catalog \citep{Sanders:2007aa} only contains sources with $S_\mathrm{24\,\micron} > 150\,\mu$Jy. We run the \texttt{SExtractor} in the S-COSMOS 24-$\micron$ image \citep{Sanders:2007aa} and recalibrated our extracted 24-$\micron$ fluxes to their \emph{Spitzer} General Observer Cycle 3 total flux. Our generated catalog has a 3.5$\sigma$ detection limit of 57\,$\mu$Jy without using positional priors from other wavelengths. We verfiy that our generated catalog is in good agreement with the deep \emph{Spitzer} catalog ($S_\mathrm{24\,\micron} > 80\,\mu$Jy) provided by \cite{LeFloch:2009aa} with a median value of $|\Delta S_\mathrm{24\,\micron}| / S_\mathrm{24\,\micron}= 6^{+10}_{-4}\%$ and with median flux peak offsets of $\Delta \alpha = 0.0\arcsec\pm0.3\arcsec$ and $\Delta \delta = 0.3\arcsec\pm0.3\arcsec$.

We utilized the multi-wavelength band-merged COSMOS2015 catalog compiled by \cite{Laigle:2016aa}, which includes 30+ bands of photometric data, spanning from X-ray through the near-ultraviolet and optical to the far-infrared. We used the combined ``FLAG\_HJMCC = 0 (or 2)" and ``FLAG\_COSMOS = 1," which is the region covered by UltraVISTA-DR2 occupying an area of 1.6\,deg$^{2}$ in the COSMOS field. The near-infrared data (e.g. $K$-band) is essential for accurate photometric redshift and stellar mass estimates, and the observed $K$-band magnitude correlates well with stellar mass up to $z\sim4$ \citep{Laigle:2016aa}. To ensure a uniform selection, we further limited ourselves to galaxies with $K$-band magnitude of $m_{\rm K}<24.5$\,mag$_{\rm AB}$ (limiting magnitude at 3$\sigma$ in a 2$\arcsec$ diameter aperture from \citealt{Laigle:2016aa}). The COSMOS2015 catalog also provides stellar mass and redshift estimations, which are used in our sample selection (\S\ref{sec:Comparison}) and clustering analysis (\S\ref{sec:Clustering}). The stellar mass and redshift measurements in the COSMOS2015 catalog were fitted using the \texttt{LE PHARE} code \citep{Arnouts:1999aa, Ilbert:2006aa}. In this work, we do not exclude AGNs from our sample and we refer readers to \cite{Laigle:2016aa} for further details.

We employed a stacking technique to assess the averaged 450- and 850-$\micron$ flux density of our SMG candidates (\S\ref{sec:Justifications}) using the wide-field SCUBA-2 image from S2COSMOS \citep{Simpson:2019aa}. S2COSMOS is an EAO Large Program that has mapped the entire COSMOS field to a uniform depth at 850 $\micron$ of $\sigma=1.2$\,mJy\,beam$^{-1}$. Thanks to the dual-band observing capability of SCUBA-2, 450-$\micron$ data were simultaneously obtained. While S2COSMOS is designed for the 850-$\micron$ imaging (so it has been carried out under weather conditions less suitable for 450-$\micron$ observations), the large area and the depth reached 1$\sigma \simeq 12$\,mJy\,beam$^{-1}$ at 450\,$\micron$ allows us to obtain strong constraints on the stacked flux of the SMG candidates.

\subsection{Training sample} \label{subsec:TrainingSample}
The methodologies employed here for source extraction and counterpart identification are similar to those in \cite{Lim:2020aa}. We briefly summarize the method here, referring readers to \cite{Lim:2020aa} for further details. 

We employed a source extraction method similar to the ``CLEAN'' deconvolution in radio interferometric imaging. We identified the most significant peak in the S/N map and subtracted 5\% of a peak-scaled model PSF from the image at its position. The subtracted flux and coordinates were then cataloged and the subtraction was iterated until a significance threshold floor ($=3.5\sigma$) was reached. We summed up the subtracted fluxes and the remaining threshold flux density and took this to be the final flux density for each source. In this work, we limited our 450-$\micron$-selected sample to the sources with total integrated S/N over the ``CLEAN"ed area $\geqslant 4$ due to the relatively high fraction of spurious sources ($\geqslant 20$\%) at S/N $< 4$. We further limited our sources to those with $S_\mathrm{450\,\micron}$ $\geqslant$ 4\,mJy\,beam$^{-1}$ to achieve a more uniform selection and so address the non-uniform sensitivity coverage of our map. In total, we obtained 221 such sources from a region of $\simeq 300$\,arcmin$^{2}$.

To construct the $K$-band source catalog for our 450-$\micron$-selected sample, we first cross-matched the positions of 450-$\micron$ sources with positions from the VLA-COSMOS 3-GHz radio catalog \citep{Smolcic:2017aa} using a 4$\arcsec$ search radius. In total, we found 131 VLA-identified sources and this procedure is expected to produce $\simeq 3$ false matches out of the 131. The expected false matches are derived from the number density of the matched catalog at a certain search radius, and multipy it by the total number of matched sources. We then cross-matched the radio positions with coordinates from the \emph{Spitzer} IRAC near-infrared catalog \citep{Sanders:2007aa} using a 1$\arcsec$ search radius ($\simeq 4$ expected false matches out of 123). For the remaining ninety 450-$\micron$ sources without radio counterparts, we cross-matched them to the \emph{Spitzer} MIPS 24-$\micron$ catalog using a search radius of 4$\arcsec$ ($\simeq 4$ expected false matches out of 65). Based on the 24-$\micron$ positions, we then made a positional matching in the IRAC near-infrared catalog by using a 2$\arcsec$ search radius ($\simeq 2$ expected false matches out of 52). Once we obtained the IRAC positions from the mid-infrared or radio counterparts, we associated these sources with the band-merged COSMOS2015 photometric catalog \citep{Laigle:2016aa} using a search radius of 1$\arcsec$. 

Among the 221 sources, 164 (74\%$\pm$6\%) of the 450-$\micron$-selected SMGs have $K$-band counterparts, of which 117 sources are VLA-identified and 47 are MIPS-identified. All of them are detected significantly according to a corrected-Poissonian probability identification technique ($p$-values $<0.05$; see \citealt{Downes:1986aa}). We employed these 164 sources as our SMG dataset for the machine-learning algorithm (\S\ref{sec:Methodology}). It is worth noting that 57 sources (26\%$\pm$3\% of total 221 sources) do not have MIPS/VLA counterparts and so they are not in the training set. This missing population does not exhibit significant variation in 450-$\micron$ flux density compared to the sources having MIPS/VLA identifications with $p$-value of 0.16 in a Kolmogorov-Smirnov test (KS test). These 57 sources most likely lie at higher redshifts ($z \gtrsim 3$; see also Figure 4 in \citealt{Lim:2020aa}) due to the fact that the mid-infrared and radio wavebands do not benefit from a strong negative K-correction, so they biased against identifying the high-redshift SMGs. Therefore, we do not expect a significant impact from this missing population on our final results (\S\ref{sec:Clustering}), which mainly focus on $z<3$. 

To construct a non-SMG sample for the training that is undetected by SCUBA-2 but within the SCUBA-2 footprint, we select 4705 $K$-band-selected sources that reside within the noise level $\leqslant 1$\,mJy\,beam$^{-1}$ region of the 450-$\micron$ image. We adopt their $K$-band magnitudes and color-color pairs (i.e., flux ratios) to be the feature vectors in the machine-learning algorithm. Given the faint optical magnitudes of most of our SMG sample, we only adopt the broad-band photometries from COSMOS2015 catalog in this work. In total, we have 79 features, of which 78 features are derived by the interlacing color quantities from thirteen-band photometry ($uBVri^{+}z^{++}JHK$[3.6][4.5][5.8][8.0]).

\section{Machine-Learning Methodology} \label{sec:Methodology}

\subsection{Performance measures of classification}\label{subsec:PerformancesMeasures}

Before introducing our machine-learning methodology, we first describe how we verify the performance of the classification. In the field of machine learning, a confusion matrix is often used to describe the performance of a classification model (Table \ref{tbl:ConfusionMatrix}). The confusion matrix has four terms: true positive (TP) refers to an actual positive sample correctly labeled as positive; false positive (FP) is a sample incorrectly flagged as positive while in reality it is negative; false negative (FN) corresponds to real positive cases incorrectly flagged as negative; and true negative (TN) represents an actual negative sample correctly labeled as negative. From these categories, one can compute the precision, recall (in other words the sensitivity or true positive rate, TPR) and false positive rate (FPR) as 
\begin{align*}
& \rm Precision = \frac{\rm TP}{\rm TP + \rm FP},  \\
& \rm Recall = \frac{\rm TP}{\rm TP + \rm FN},\,and \\
& \rm FPR = \frac{\rm FP}{\rm FP + \rm TN}. 
\end{align*}
The precision represents the proportion of all correct positive predictions, while recall is the recovery of all real positive cases that are predicted to be positive. Conversely, the FPR is the ratio between the number of actual negative cases wrongly categorized as positive and the total negative cases. 

Several meaningful indicators are often used to verify the performance of a classifier. The f1-measure \citep{Rijsbergen:1979aa} considers both precision and recall to compute a score. This f1-score can be interpreted as a harmonic average of precision and recall, which can be measured as 
\begin{equation}
\rm f1\textnormal{-}score = \frac{2 \times \rm Precision \times \rm Recall}{\rm Precision + \rm Recall}.
\end{equation}
The f1-score reaches its best value at 1 and worst at 0. 

Another standard test for evaluating a binary decision problem is using the Receiver Operating Characteristic (ROC) Curve that is plotting TPR against FPR \citep{Provost:1998aa}. A perfect classifier, which has no FN and FP, will have a high value of TPR and low value of FPR. Therefore, a higher value of the area under the ROC curve (AUROCC) corresponds to a better classifier. 

	\begin{deluxetable}{ccc}
	\tablecaption{\label{tbl:ConfusionMatrix} Confusion matrix for binary classification. }
	\tablehead{  \colhead{  } & \colhead{ Positive prediction } &  \colhead{ Negative prediction } }
	\startdata
	Positive class  & True Positive (TP)  & False Negative (FN) \\
	Negative class & False Positive (FP) & True Negative (TN) \\
	\enddata
	\end{deluxetable}

\subsection{Methodology} \label{subsec:Methodology}

In this work, we adopt the extreme gradient boosting (XGBoost, \citealt{Chen:2016aa}) method that is designed based on a scalable gradient tree boosting learning, since the XGBoost performs the best for identifying the SMG candidates in our sample (\S\ref{subsec:AlgorithmsComparison}; see also other similar works done with XGBoost, \citealt{An:2018aa, An:2019aa, Liu:2019aa}). The idea of gradient tree boosting is to build an ensemble of simpler estimators (usually they are decision trees) and convert them sequentially into a complex predictor. During the iterations of building the trees, each tree will correct the error between the predictions and the actual output from the existing trees and minimize the training error of the ensemble. The contribution of each tree can be scaled to reduce the influence of each tree that will leave more space for future trees to improve the ensemble. This process will lead to a better model and prevent the behavior of over-fitting \citep{Friedman:2000, Friedman:2002}. Typically, smaller values of this weighting (i.e., learning rate or shrinkage) seem to produce a better performance \citep{Friedman:2000}. 

XGBoost is designed to push the limit of computational resources and improve the model performance for the gradient tree boosting algorithm. XGBoost performs a split finding algorithm that enumerated over all possible splits on all features and finds the best split in tree learning. The advantage of XGBoost over other techniques is that it can handle missing features. When the algorithm encounters a missing value, it is labeled into the default direction which is already learned from the data. 

In this work, we define 70\% of our data to be the training set and the rest as the test set, where the training set builds the model and the testing set evaluates the model's performance. In principle, the predictor performs better with a larger fraction of training data. On the other hand, the performance statistic will have greater variance if there is less testing data \citep{Kohavi:1995aa}. To strike a balance, we have tried changing the ratios of our training and testing sets from 50:50 to 90:10, we verify that the XGBoost performs the best in both AUROCC and f1-score with a 70:30 split.

To avoid over-fitting, we adopt an early stopping in XGBoost after five training iterations that do not yield any improvements. To control the balance of positive and negative weights, we set the $scale\_pos\_weight$ = 28.7, which is given by the ratio between the number of negative and positive instances (4705/164, see \S\ref{subsec:TrainingSample}).

\subsection{Feature selection} \label{subsec:FeatureSelection}
Feature selection is an important process in machine learning that strongly influences the performance of the model. Feature selection is a procedure of selecting a subset of relevant features for model construction. In general, proper feature selection can increase the efficiency by reducing the training duration, enhance generalization by reducing overfitting, and improve the prediction performance (see \citealt{Chen:2016aa, An:2019aa}). 

A trained XGBoost model will automatically calculate feature importance and provide the feature importance scores. In this work, we first trained the XGBoost model based on the training dataset and selected the features by sorting the feature importance scores calculated from the trained model. We then iteratively trained the model based on the selected subset of features until the point of best performance. We verified that both f1-score and AUROCC increase with the number of selected features until we use up all the 79 features. Therefore, we did not reduce the number of feature vectors in this work. Considering the limitation of our computational resources, we repeated this procedure ten times by using a different combination of training and test datasets in each iteration and estimated the average feature importance score from these ten realizations. The top five important features in our sample are $K$, ([3.6]$-$[4.5]), ($K$$-$[4.5]), ([3.6]$-$[5.8]), and ($H$$-$$K$), which are similar to those photometric wavebands used by \citealt{Chen:2016ab} (Optical-Infrared Triple Color: ($z$$-$$K$), ($K$$-$[3.6]), and ([3.6]$-$[4.5])) and \citealt{An:2019aa} (($z$$-$$K$), ($J$$-$$K$), ($K$$-$[3.6]), and ([3.6]$-$[4.5])). The top five important features in our sample can be associated to fundamental physical properties. The $K$ band flux roughly maps to stellar mass, while the SMGs appear to be red and occupy a relatively well-defined region in near-infrared color-color space (see also \citealt{Chen:2016ab, An:2019aa}).

\subsection{Tuning the hyper-parameters} \label{subsec:TuningParameters}
The hyper-parameters are a set of parameters that define the machine-learning algorithm as a mathematical formula. The hyper-parameters act as tuning functions that are set during the training of the model. We optimized the hyper-parameters in XGBoost using $k$-fold cross-validation \citep{An:2019aa}. This is a resampling procedure used to iteratively evaluate the performance of machine-learning models on a limited data sample. In each iteration, this procedure randomly divides the training dataset into $k$ groups/folds (approximately equal size). Each unique fold is treated as a validation set and the model is fitted on the remaining $k-1$ folds. The validation set is replaced $k$ times and the average performance measure of the $k$ sets is then reported. In this work, we used $k=5$ and adopted the AUROCC as the scoring function to optimize the hyper-parameters of the XGBoost classifier. We used the \texttt{RandomizedSearchCV} recipe from the \texttt{python}-based \texttt{scikit-learn} package \citep{Pedregosa:2011aa}. \texttt{RandomizedSearchCV} randomly searches over a combination of hyper-parameter space and finds the best solution for the constructed model. In the same computational budget, \texttt{RandomizedSearchCV} can find models that are as good or better compared to a pure grid search \citep{Bergstra:2012aa}. In this work, we set the number of iterations to 200. We verify that both f1-score and AUROCC converge after 200 iterations (confirming up to 300 iterations).

\subsection{Algorithms comparison} \label{subsec:AlgorithmsComparison}
We repeated the procedure in \S\ref{subsec:TuningParameters} ten times by using the exactly same combinations of training and test datasets used in \S\ref{subsec:FeatureSelection} in each iteration. Each optimizer gives different results in labeling the SMG candidates in accordance with the expectation. We verify that the labeled SMG candidates change within 10\% from each optimizer. The mean performance from these ten optimizers is summarized in Table \ref{tbl:AlgorithmsComparison}. 

We also test the performance of XGBoost against other machine-learning algorithms. To make a fair comparison, we replaced the missing data with the detection limits of each feature vector since most of the other algorithms do not handle missing values. We optimized each of the algorithms (similar to what was done in \S\ref{subsec:TuningParameters}) and repeated the procedure ten times by using the exactly same combinations of training and test datasets used in \S\ref{subsec:FeatureSelection} in each iteration. As shown in Table \ref{tbl:AlgorithmsComparison}, the XGBoost method with missing values performs the best in terms of both AUROCC and f1-score. Furthermore, we verify that the values of AUROCC, f1-score, precision and recall do not change significantly with redshift when we review our test dataset in specific redshift bins. 

\subsection{Result from the XGBoost algorithm} \label{subsec:XGBResult}
To accommodate the different results from each optimizer due to different combinations of training and test datasets, we adopt the following approach so that the training results are close to the mean performance from the ten optimizers mentioned in \S\ref{subsec:TuningParameters}. We combined the predicted class probabilities (output from \texttt{predict\_proba} algorithm in \texttt{scikit-learn} package) from each optimizer by using a combined probability formula $(P_{1}P_{2}...P_{n})^{\frac{1}{n}}$, where $n$ is 10 in this work. We labeled the sources that have a final class probability $\geqslant 0.5$ (the default threshold for two-classes classification in XGBoost), as SMG candidates. This procedure is similar to the bootstrap aggregating, so called bagging \citep{Breiman:1996aa}. Bagging is a two-step process: bootstrapping and aggregating. Bootstrapping is a sampling method that randomly select several subsets of samples from the entire dataset. The individual subset of samples is then taken as the training dataset for the machine-learning models. The aggregation then combines the model predictions from those subsets into a final prediction considering all the outcomes possible. In short, bagging procedure can generate an aggregated predictor from multiple predictors, which can improve the stability and accuracy of machine-learning algorithms and reduce variance to avoid overfitting \citep{Breiman:1996aa}. 

To validate the performance of the aggregated predictor, we isolated 30\% of our sample as the independent test sample and split the remaining 70\% into 50\%:20\% for training. We trained the 50\%+20\% dataset ten times in XGBoost, in which the 50\% and 20\% subsamples were randomly drawn in each iteration.. We used the trained classifiers to estimate the predicted class probabilities of the isolated 30\% test set in each iteration and did the bagging procedure in their probabilities. We confirmed that the performance of the aggregated predictor (precision and recall) is almost the same as the mean performance (precision and recall) from those ten iterations, indicating that the aggregated predictor is a representative classifier for a number of classifiers. Therefore, we conclude that the performance of the aggregated classifier (after the bagging procedure) should be similar to the mean performance shown in Table \ref{tbl:AlgorithmsComparison}.

Based on the adopted training results there is a non-negligible degree of misidentification (precision $=0.59\pm0.04$) and incompleteness (recall $=0.68\pm0.07$) in our classifications. It is therefore worth investigating what population is labeled incorrectly as SMGs while in reality it is not (i.e., the FP) and what fraction of SMGs are labeled incorrectly as field galaxies (i.e., the FN) by our trained XGBoost algorithm. To do this, we investigate the properties of FP in each of the ten iterations. The median stellar mass and median redshift of our training 450-$\micron$ SMGs are log($M_{\ast}/\rm M_{\sun}$) = $10.76^{+0.04}_{-0.02}$ and $z = 1.64^{+0.16}_{-0.07}$, respectively, and their median stacked flux is ($7.1\pm0.5$)\,mJy at 450\,$\micron$ based on our STUDIES image. We find that the overall properties of FP have a median redshift of $z=2.0\pm0.3$, median stellar mass of log($M_{\ast}/\rm M_{\sun}$) = $10.7\pm0.1$, and median stacked flux of ($2.5\pm0.5$)\,mJy at 450\,$\micron$, based on our STUDIES image. These findings show that the FP has similar stellar mass to the training SMG sample, but appear to have slightly higher redshifts compared to the training SMG sample and are less active in obscured star formation. Meanwhile, we carry out the same investigation for the FN. The overall properties of FN have median redshift $z=1.4\pm0.2$, median stellar mass log($M_{\ast}/\rm M_{\sun}$) = $10.5\pm0.1$, and median stacked flux ($6.0\pm0.5$)\,mJy at 450\,$\micron$. These findings show that our classifier may miss the population that is slightly less massive and lower in redshift compared to our training SMG sample. We conclude that the effectiveness of our classifier could be hindered by the fuzzy boundary between the SMGs and the less-active star-forming galaxies having similar stellar masses. We do not expect that this finding will impact our final clustering measurements significantly in \S\ref{sec:Clustering}, where we show that the clustering signals of star-forming galaxies are very similar to the SMG candidates once they are matched in redshift and stellar mass.

As an additional test in Appendix \ref{sec:ClusteringResultsOther} we have applied two distinct machine learning algorithms with different levels of recall and precision and show that the clustering signals recovered for the SMGs are robust. The SMG candidates that are identified by decision tree (better recall) and random forest (better precision) yield clustering signals which are not significantly different from the results of XGBoost identified SMG candidates, indicating that our final results are insensitive to the chosen classifiers (see \S\ref{sec:ClusteringResultsOther}).

The spurious sources in the SMG training sample, which are wrongly labeled as SMGs, may impact the training sample and bias the performance estimators. We check that the cumulative spurious fraction of our SMG training sample is $\simeq10$\%. To quantify how sensitive our training is to this contamination, we randomly choose 10\% of the SMG training sample and swap them with the MIPS- or VLA-detected non-SMGs in the STUDIES field. We then train the XGBoost by using this training sample and repeat the procedure ten times. The means AUROCC and f1-scores from these procedures are reduced by 3\% and 10\%, respectively, compared to the measurements shown in Table \ref{tbl:AlgorithmsComparison}. These offsets are expected, since the contamination in the training sample will reduce the precision of the predictor. Considering the uncertainties in the measured AUROCC and f1-score, we conclude that our final clustering measurements (\S\ref{sec:Clustering}) will not be impacted significantly by the small fraction of spurious sources.

In total, our trained XGBoost algorithm labels 6182 SMG candidates from the COSMOS2015 catalog containing 307374 sources that have an $m_{\rm K}<24.5$\,mag$_{\rm AB}$ across an effective area of 1.6\,deg$^{2}$.

\begin{deluxetable*}{ccccc}
\tablecaption{\label{tbl:AlgorithmsComparison} Performance of test samples using several machine-learning methods. We replace the missing data with the detection limits of each feature vector, since most of the other algorithms do not handle missing values. The XGBoost method with missing values (marked in bold) performs the best for both AUROCC and f1-score.}
\tablehead{  \colhead{ Methodology } & \colhead{ AUROCC } &  \colhead{ f1-score } &  \colhead{ Precision } &  \colhead{ Recall } } 
\startdata
Adaptive boosting                        & $0.97\pm0.01$ & $0.55\pm0.07$ & $0.80\pm0.07$ & $0.43\pm0.09$ \\
Decision tree           			 & $0.92\pm0.02$ & $0.36\pm0.03$ & $0.23\pm0.03$ & $0.87\pm0.07$ \\
Logistic regression 			 & $0.97\pm0.01$ & $0.56\pm0.07$ & $0.77\pm0.08$ & $0.46\pm0.10$ \\
Random forest        			 & $0.97\pm0.01$ & $0.54\pm0.07$ & $0.81\pm0.08$ & $0.41\pm0.07$ \\
Stochastic gradient boosting     & $0.96\pm0.01$ & $0.61\pm0.03$ & $0.74\pm0.06$ & $0.53\pm0.05$ \\
Support vector machines           & $0.91\pm0.04$ & $0.51\pm0.04$ & $0.78\pm0.06$ & $0.38\pm0.05$ \\
XGBoost 				                   & 0.97$\pm$0.01 & 0.57$\pm$0.07 & $0.51\pm0.14$ & $0.73\pm0.12$ \\
\textbf{XGBoost (with missing values)} & \textbf{0.97}$\pm$\textbf{0.01} & \textbf{0.63}$\pm$\textbf{0.02} & \textbf{0.59}$\pm$\textbf{0.04} & \textbf{0.68}$\pm$\textbf{0.07} \\
\enddata
\end{deluxetable*}

\section{Verifications of the SMG candidates} \label{sec:Justifications}

	\begin{figure*}
	\gridline{ \fig{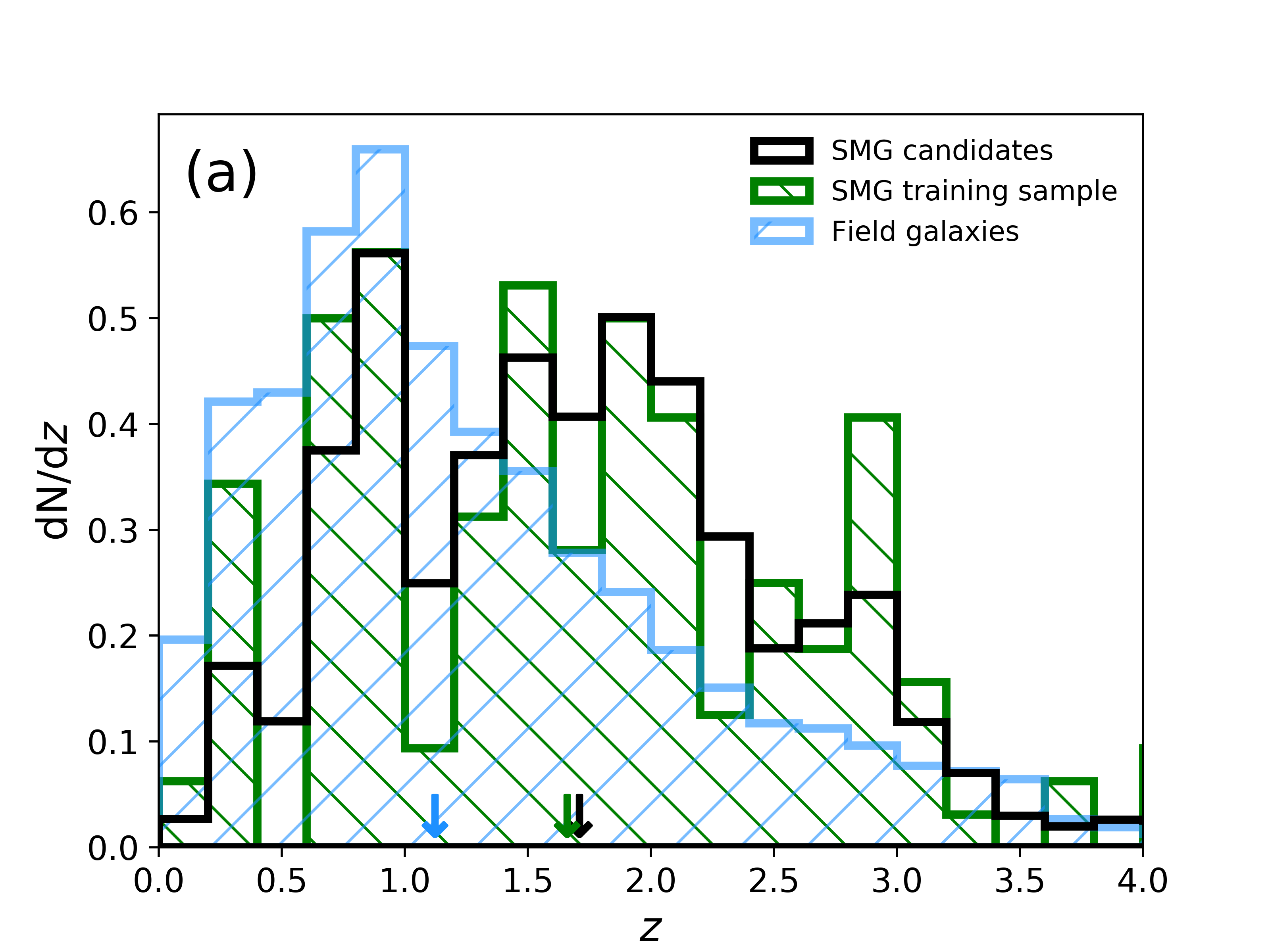}{0.5\textwidth}{}\hspace{-2em}
	\fig{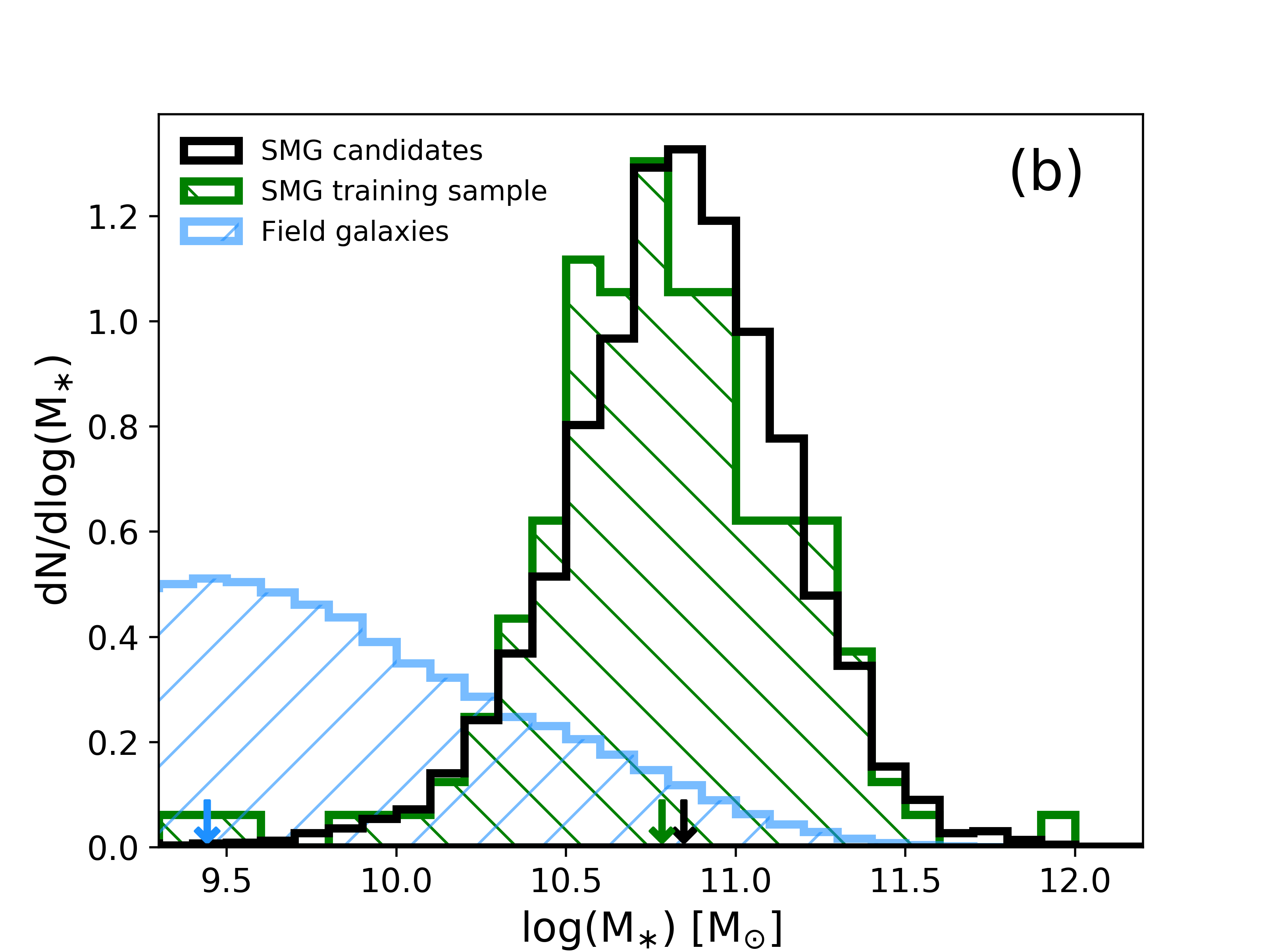}{0.5\textwidth}{}\hspace{-2em}}
	\caption{Panel (a): normalized histograms of photometric redshift; and panel (b): normalized histograms of stellar mass for the 450-$\micron$-selected SMG training sample, SMG candidates, and field galaxies. The median values are marked as downward arrows for the corresponding sample. The median redshifts and stellar masses of SMG candidates are consistent with the estimations from the parent SMG training dataset.}
	\label{fig:TrainSample}
	\end{figure*}

	\begin{figure}
	\includegraphics[width=\columnwidth]{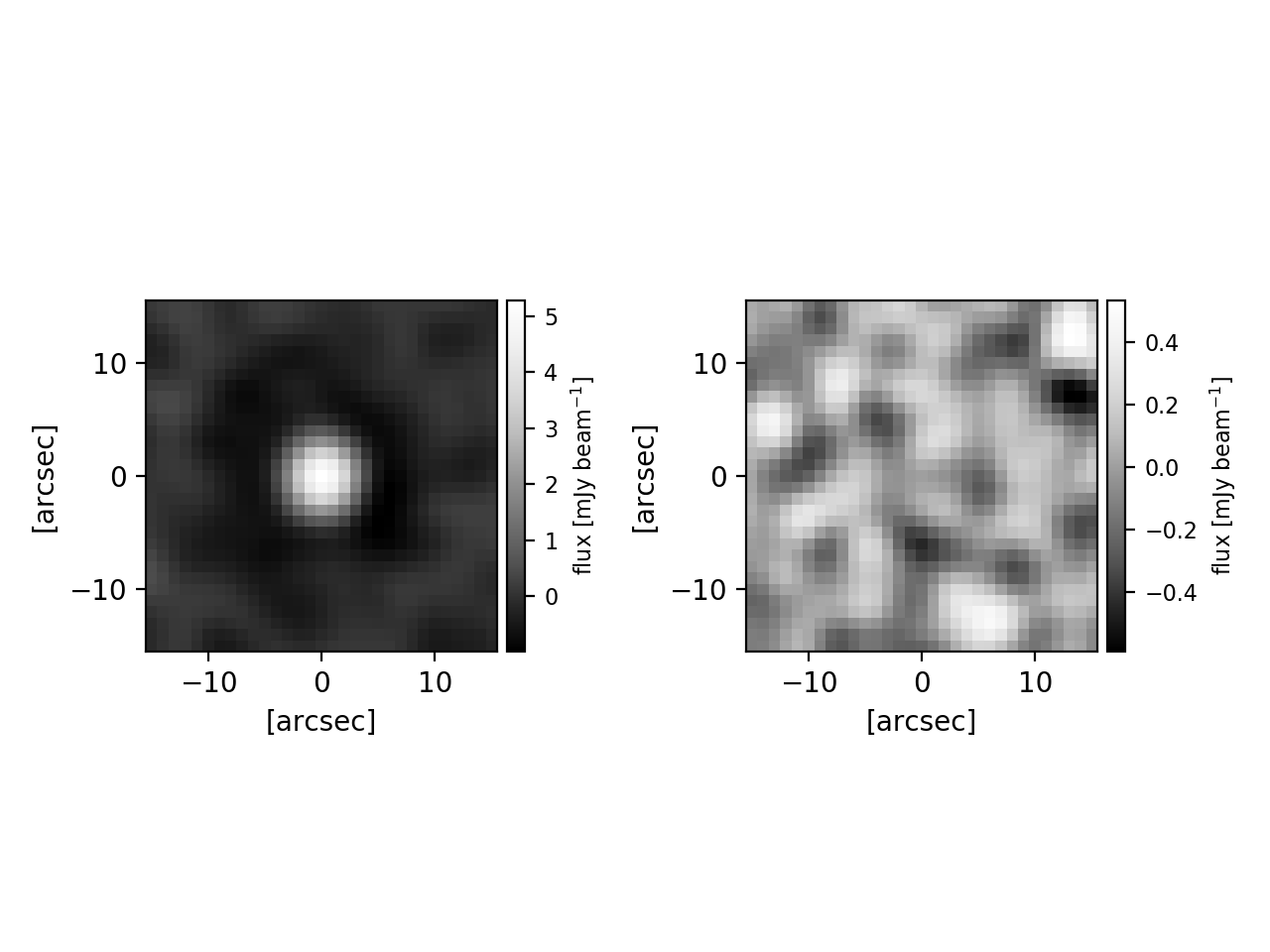}
	\caption{Left: stacked SCUBA-2 450-$\micron$ image from the S2COSMOS survey \citep{Simpson:2019aa} at the positions of 6182 SMG candidates. The negative flux trough surrounding the source is caused by the matched-filter procedure from the \texttt{SCUBA2\_MATCHED\_FILTER} recipe (see \S\ref{sec:Data}). The stacked 450-$\micron$ flux density is ($5.3\pm0.2$)\,mJy. Right: stacked 450-$\micron$ image at 6182 random positions.}
	\label{fig:StackedImage_SMGCandidates}
	\end{figure}

While machine learning is a powerful and promising technique for data analysis, it inevitably appears to be a black box to typical users. Checks need to be performed to ensure that the selected SMG candidates have properties (especially those not used for the training) similar to those of the parent 450-$\micron$-selected SMG sample.

The total number of SMG candidates (6182 sources) is consistent with the predicted number counts of 6100$^{+1800}_{-1400}$ of 450-$\micron$ sources in the field. The predicted number count is derived by integrating the best-fitted function of \cite{Wang:2017aa} over a flux range of 3.4\,mJy to 36\,mJy. The predicted median flux in this flux density range is ($6.8\pm0.1$)\,mJy. The lower end of the range is determined from the boosting corrected 450-$\micron$ flux density range of our training SMG dataset. We further apply the fraction of $0.74\pm0.06$ to take into account the selection bias in our training dataset (see \S\ref{subsec:TrainingSample}), since our training SMG dataset only includes SMGs having $K$-band counterparts. 

To probe the sub-millimeter emissions of our 6182 SMG candidates, we directly measure their 450-$\micron$ flux density from the S2COSMOS image. The 450-$\micron$ flux distribution of SMG candidates is consistent with the measurement from the 450-$\micron$-selected SMG training sample, based on the same image ($p$-value $= 0.36$ in KS test). The median stacked 450-$\micron$ flux density from the S2COSMOS image of the SMG candidates is ($5.3\pm0.2$)\,mJy (Figure \ref{fig:StackedImage_SMGCandidates}). This value is tentatively offset ($\simeq 2.3\sigma$) from the median stacked flux of the 450-$\micron$-selected SMG training sample ($7.8\pm1.1$)\,mJy derived from the same S2COSMOS image but significantly offset from the aforementioned predicted median flux of ($6.8\pm0.1$)\,mJy and the median stacked flux of the SMG training sample, based on our deeper STUDIES image ($7.1\pm0.5$)\,mJy. The stacked value of the SMG training sample could be biased high, since sources close to the detection threshold will be included if they are on a peak of the noise. Indeed, the deboosted 450-$\micron$ stacked flux of the 450-$\micron$-selected SMG training sample is $6.4\pm0.5$\,mJy, only marginally higher than the stacked flux of the SMG candidates. To have an independent test, we apply the stacking analyses in the 850-$\micron$ images from the S2COSMOS and STUDIES surveys by assuming that the 450- and 850-$\micron$ noise maps are reasonably independent. The median stacked 850-$\micron$ flux density from the S2COSMOS image of our SMG candidates is ($1.23\pm0.03$)\,mJy. This value is significantly offset from the median stacked 850-$\micron$ flux of the 450-$\micron$-selected SMG training sample ($1.9\pm0.2$)\,mJy based on the same image, but tentatively offset ($\simeq 1.9\sigma$) from similar measurements based on our deeper STUDIES image ($1.7\pm0.2$)\,mJy. The fainter stacked flux of the SMG candidates compared to the stacked flux of SMG training sample is expected, since our machine-learning algorithm misidentifies a fraction of SMGs ($\simeq 30\%$; see the recall value in Table~\ref{tbl:AlgorithmsComparison}). However, the stacked flux only provides the overall emission properties of our sample. We do not expect that our results in the following analyses will be affected significantly due to the contamination, since the contamination arises from less dusty star-forming galaxies with similar stellar masses to the training SMG sample (see \S\ref{subsec:XGBResult}) and the clustering signals of star-forming galaxies are very similar to the SMG candidates once they are matched in redshift and stellar mass (\S\ref{sec:Clustering}).

All of the 450-$\micron$-selected SMG training sample in the machine-learning algorithm has matched detections from the MIPS at 24\,$\micron$ or VLA at 3\,GHz, or both (see \S\ref{subsec:TrainingSample}). The fraction of our SMG candidates that have MIPS and/or VLA detections is $\simeq88\%$\,$\pm$\,1\% (=5442/6182), indicating that our SMG candidates behave similarly at mid-infrared and/or radio wavelengths compared with the SMG training sample. In addition, the MIPS- and VLA-flux distributions of our SMG candidates are consistent with the SMG training dataset, having $p$-values of 0.77 and 0.31 in the KS test, respectively. Based on the S2COSMOS image, the stacked flux of our SMG candidates with MIPS or VLA detections is ($5.6\pm0.2$)\,mJy, while the stacked flux of the remaining SMG candidates is ($3.0\pm0.5$)\,mJy. This indicates that $\simeq12$\% (=740/6182) of our SMG candidates are misidentified and/or are biased toward fainter SMG population. Again, we do not expect that our results in the following analyses will be affected significantly due to this finding. It is worth noting that the MIPS 24-$\micron$ and VLA 3-GHz photometry are not part of the training features (\S\ref{subsec:FeatureSelection}). The high MIPS- or VLA-detection rates strongly support the reliability of our machine-learning algorithm. 

By comparing other populations extracted from the COSMOS2015 catalog, we can test whether the distribution of our SMG candidates is consistent with the 450-$\micron$-selected SMG training sample. Figure \ref{fig:TrainSample} shows the normalized histograms of photometric redshift and stellar mass for the 450-$\micron$-selected SMG training sample, SMG candidates, and field galaxies. The median redshift of our SMG candidates is $z = 1.69\pm0.02$ (Figure \ref{fig:TrainSample}(a)), which is in excellent agreement with that of the SMG training sample ($z = 1.64^{+0.16}_{-0.07}$). The redshift distribution of our SMG candidates is consistent with the SMG training dataset, having a $p$-value $=0.63$ in the KS test, indicating that we cannot reject the null hypothesis of no difference between our SMG candidates and parent SMG training sample. On the other hand, the $p$-value is essentially zero in the KS test between the redshift distribution of our SMG candidates and that of the field galaxies in the COSMOS2015 catalog. 

Figure \ref{fig:TrainSample}(b) shows the stellar-mass distributions of the SMG candidates, 450-$\micron$-selected SMG training sample, and the field galaxies over a redshift range of $z=0$--6. The median stellar mass of our SMG candidates (log($M_{\ast}/M_{\sun}$) = $10.83\pm0.01$) is consistent with the median stellar mass of the SMG training sample (log($M_{\ast}/M_{\sun}$) = $10.76^{+0.04}_{-0.02}$). A KS test performed on the stellar-mass distributions of SMG candidates and parent SMG training dataset shows that the test cannot distinguish between these two populations at the 95\% significance level ($p$-value $=0.15$). The KS test rejects the null hypothesis that the stellar-mass distributions of our SMG candidates and field galaxies are drawn from the same distribution ($p$-value $\simeq 0$). The similarities of physical properties (e.g. redshift and stellar mass) between our SMG candidates and the parent SMG training dataset are expected. This is because the features for estimating the physical properties in the COSMOS2015 catalog, which are based on the photometric data, are similar to the adopted features in our machine-learning algorithm.

We also test our SMG-identification technique on ALMA observations. ALMA follow-up observations resolved 260 850-$\micron$-selected SMGs in the S2CLS-COSMOS program (AS2COSMOS; Simpson et al. 2020, in preparation) from 183 850-$\micron$ sub-millimeter sources with S/N $>$ 4.3$\sigma$ (rms $\simeq 0.2$\,mJy). There are 165 ALMA-detected sources with $K$-band counterparts in the COSMOS2015 that further satisfy our selection criterion of $m_{\rm K}<24.5$\,mag$_{\rm AB}$ (see \S\ref{subsec:AncillaryData}). Among these, our machine-learning algorithm successfully identifies 126 ALMA-detected sources as SMG candidates, indicating that the completeness of our identification is 76\%\,$\pm$\,7\%. On the other hand, 148 SMG candidates are located within the 183 ALMA primary-beam areas (FWHM = 17${\farcs}$3). Among them, 121 sources are detected by ALMA, suggesting that the precision of our machine-learning classifier is 82\%\,$\pm$\,7\%. However, the ALMA observations should not necessarily have high identification rates in our sample, since ALMA and JCMT were observing in different wavelengths, and therefore, the sensitivity limits are different. To coarsely estimate the 450-$\micron$ sensitivity in the AS2COSMOS survey, we employ the typical $S_\mathrm{450\,\micron}$/$S_\mathrm{850\,\micron}$ ratio of 2.5--4.5 at the faint end \citep{Hsu:2016aa}. The sensitivity of AS2COSMOS ($\simeq 0.86$\,mJy) is equivalent to a 450-$\micron$ sensitivity of $\simeq 2$--4\,mJy, which is roughly our selection limit. Therefore, we conclude that our SMG candidates should be detected with AS2COSMOS and the high identification rate of AS2COSMOS in our sample is expected, which again supports the reliability of our machine-learning algorithm.

\section{Comparison samples} \label{sec:Comparison}
To put the SMG candidates into the context of general galaxy populations at the same redshifts, we construct comparison samples of star-forming galaxies and passive galaxies.

	\begin{figure*}
	\gridline{ \fig{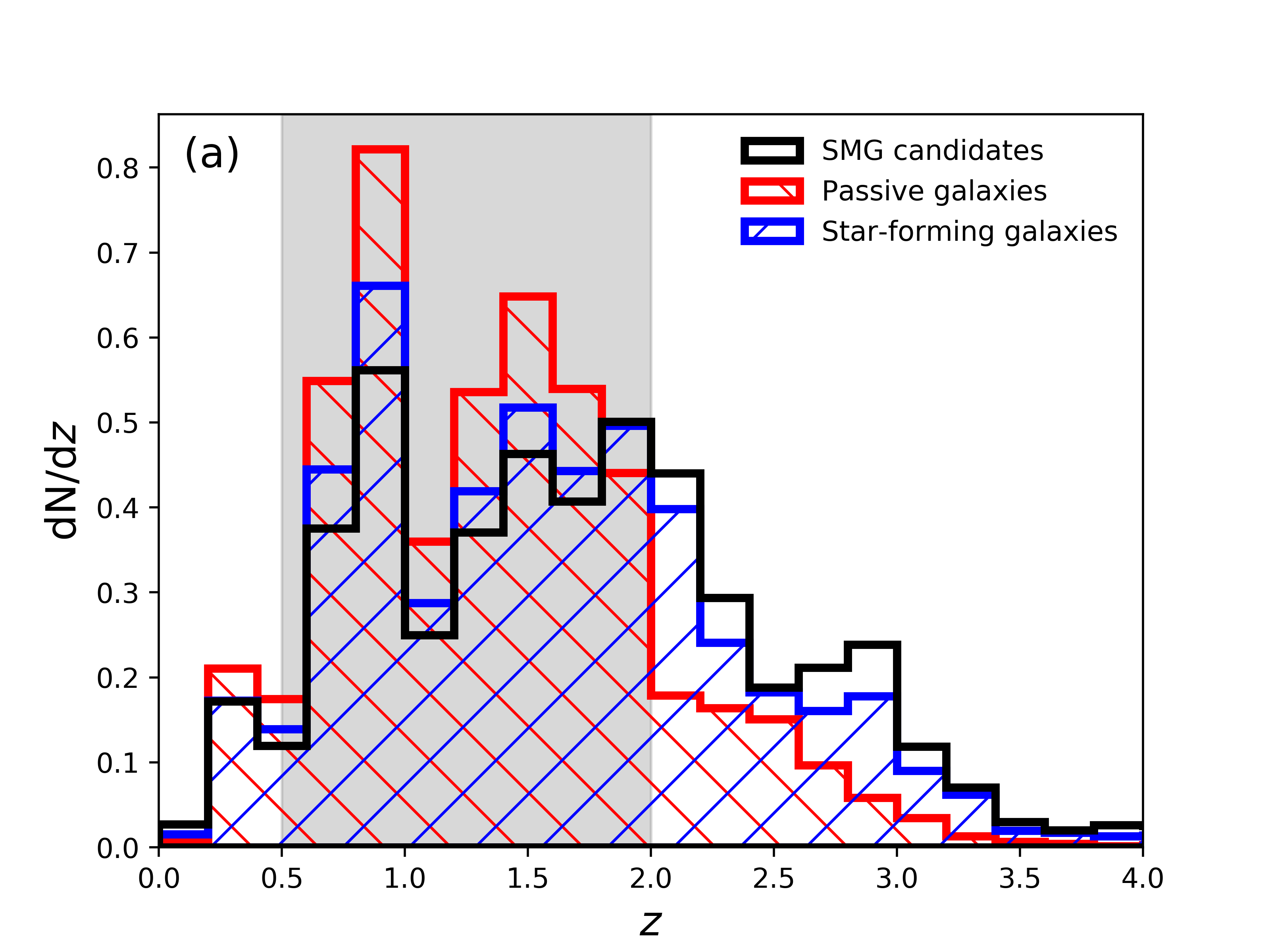}{0.5\textwidth}{}\hspace{-2em}
	\fig{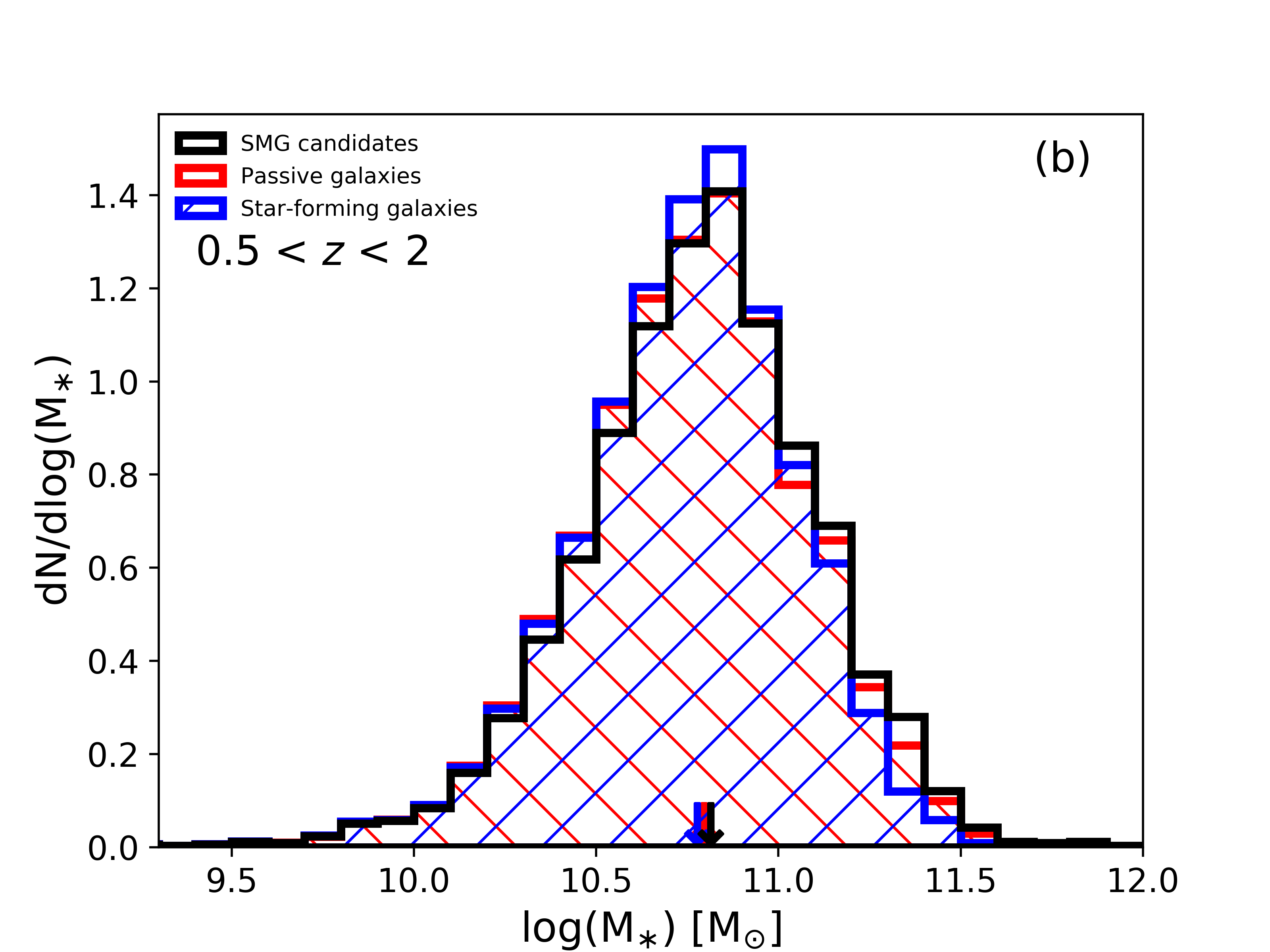}{0.5\textwidth}{}\hspace{-2em}}
	\caption{(a) Normalized redshift distributions for SMG candidates, and comparison samples of star-forming galaxies and passive galaxies. All these populations are matched as closely as possible in redshift, stellar mass, and sample size. In this work, we restrict the comparison samples to the galaxies at $0.5<z<2$ (gray-shaded region), since it is difficult to find sufficient numbers of passive and massive star-forming galaxies at $z>2$. (b) Normalized stellar-mass distributions for SMG candidates, and comparison samples of star-forming galaxies and passive galaxies at $0.5<z<2$.}
	\label{fig:PropertiesDistribution}
	\end{figure*}

\subsection{Rest-frame NUV-r-J color} \label{subsec:Selection}

Our comparison star-forming and passive galaxies are selected based on the rest-frame $M_{\rm NUV}-M_{\rm r}$ and $M_{\rm r}-M_{\rm J}$ color cuts \citep{Ilbert:2013aa} for galaxies not flagged as SMG candidates by our machine-learning algorithm. To avoid incompleteness, in the following analyses, we only consider sources that have stellar-mass estimates above the 95\% mass completeness. To empirically estimate the stellar mass completeness as a function of redshift, we follow a procedure similar to that in \cite{Pozzetti:2010aa}. We take the 20\% faintest galaxies in $K$-band magnitude in several redshift bins to be a representative observational limit for our whole sample. We then find the 95th upper percentile from the stellar mass distribution of this sub-sample in each of the redshift bins and take the values to be the stellar-mass limit for the corresponding redshift bins. The 95\% mass completeness as a function of redshift can be described with a polynomial function log$(M_{\rm lim}/{\rm M_{\sun}}) = 8.14 + 0.95z - 0.09z^{2}$.

Since it is known that galaxy clustering evolves with redshift and is a strong function of stellar mas \citep{McCracken:2015aa}, to make a fair comparison, we need to construct a sample of comparison galaxies that are matched as closely as possible to our SMG candidates in redshift, stellar mass, and sample size. We first adopt the \texttt{binned\_statistic\_2d} algorithm from \texttt{scipy} package \citep{Jones:2001aa} to generate the two-dimensional histograms of our SMG candidates and comparison galaxies by using specific redshift and stellar mass bins. We then randomly select a number of comparison galaxies in each bin, which is matched with the SMG candidates. The normalized redshift distributions for our SMG candidates and comparison samples are shown in Figure \ref{fig:PropertiesDistribution}(a). As we can see, the number of passive galaxies drops at $z>2$. Similarly, it is also hard to find sufficient numbers of massive star-forming galaxies that are not SMG candidates at $z>2$. Therefore, in this work, we restrict the comparison samples to the galaxies at $0.5 < z < 2$. In total, we randomly select 3021 and 3083 passive galaxies and star-forming galaxies, respectively, at $0.5 < z < 2$ (summarized in Table \ref{tbl:Clustering}). The redshift estimations for SMG candidates may be less reliable compared to the comparison passive and star-forming galaxies, since SMG candidates are expected to be dusty and consequently fainter in the optical. Interestingly, at $0.5 < z < 2$, the median redshift errors for SMG candidates ($\sigma_{z} = 0.02\pm0.06$) is statistically consistent with the comparison passive ($\sigma_{z} = 0.02\pm0.06$) and star-forming galaxies ($\sigma_{z} = 0.03\pm0.06$). Therefore, we conclude that the redshift estimations for all our samples are equally reliable. The stellar mass distribution for sources at $0.5 < z < 2$ is plotted in Figure \ref{fig:PropertiesDistribution}(b). The samples all show similar stellar-mass distributions according to the KS test ($p$-value$>0.05$).

	\begin{figure*}
	\gridline{ \fig{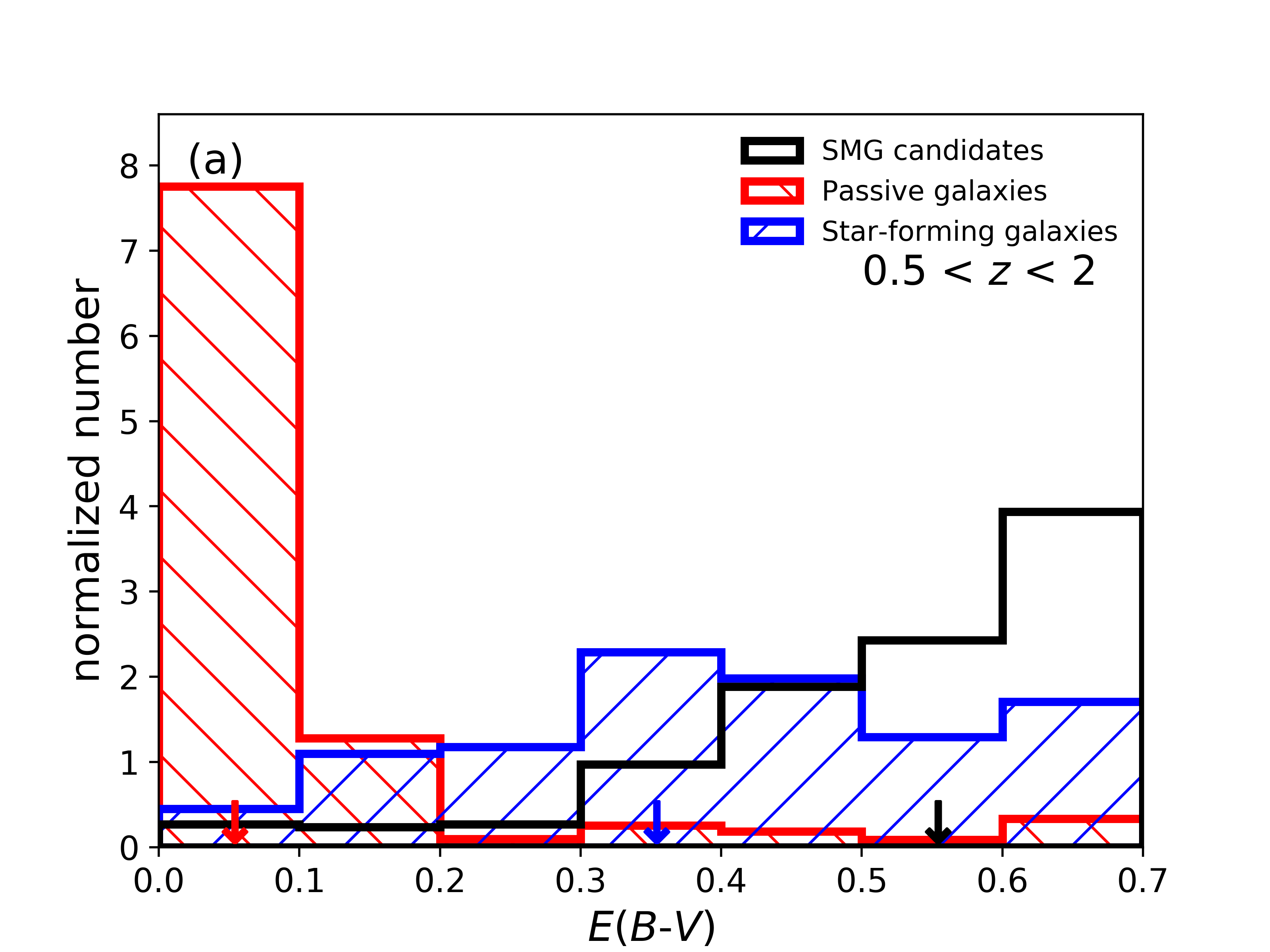}{0.5\textwidth}{}\hspace{-2em}
	\fig{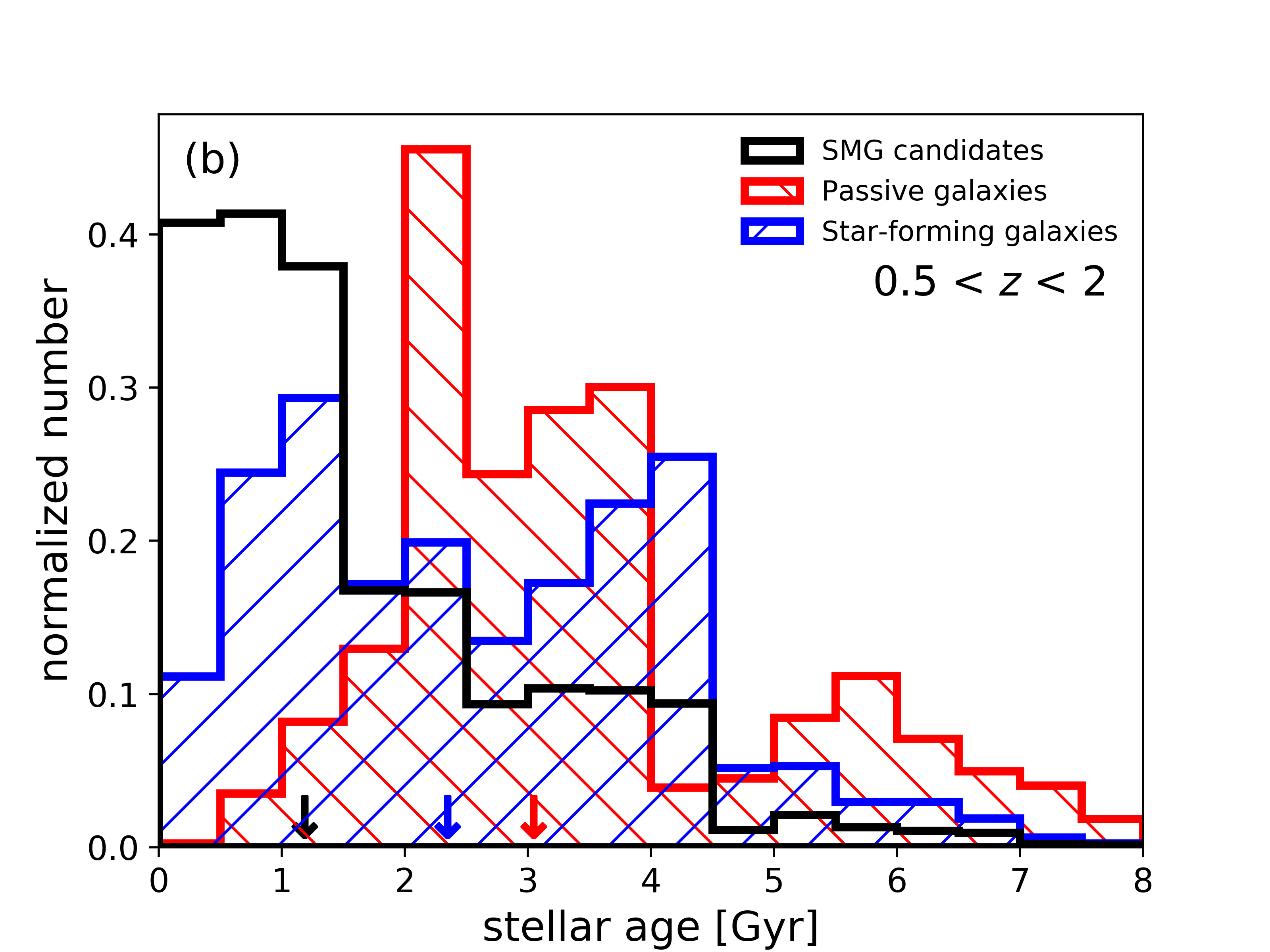}{0.5\textwidth}{}\hspace{-2em}}
	\vspace{-2.5em}
	\gridline{ \fig{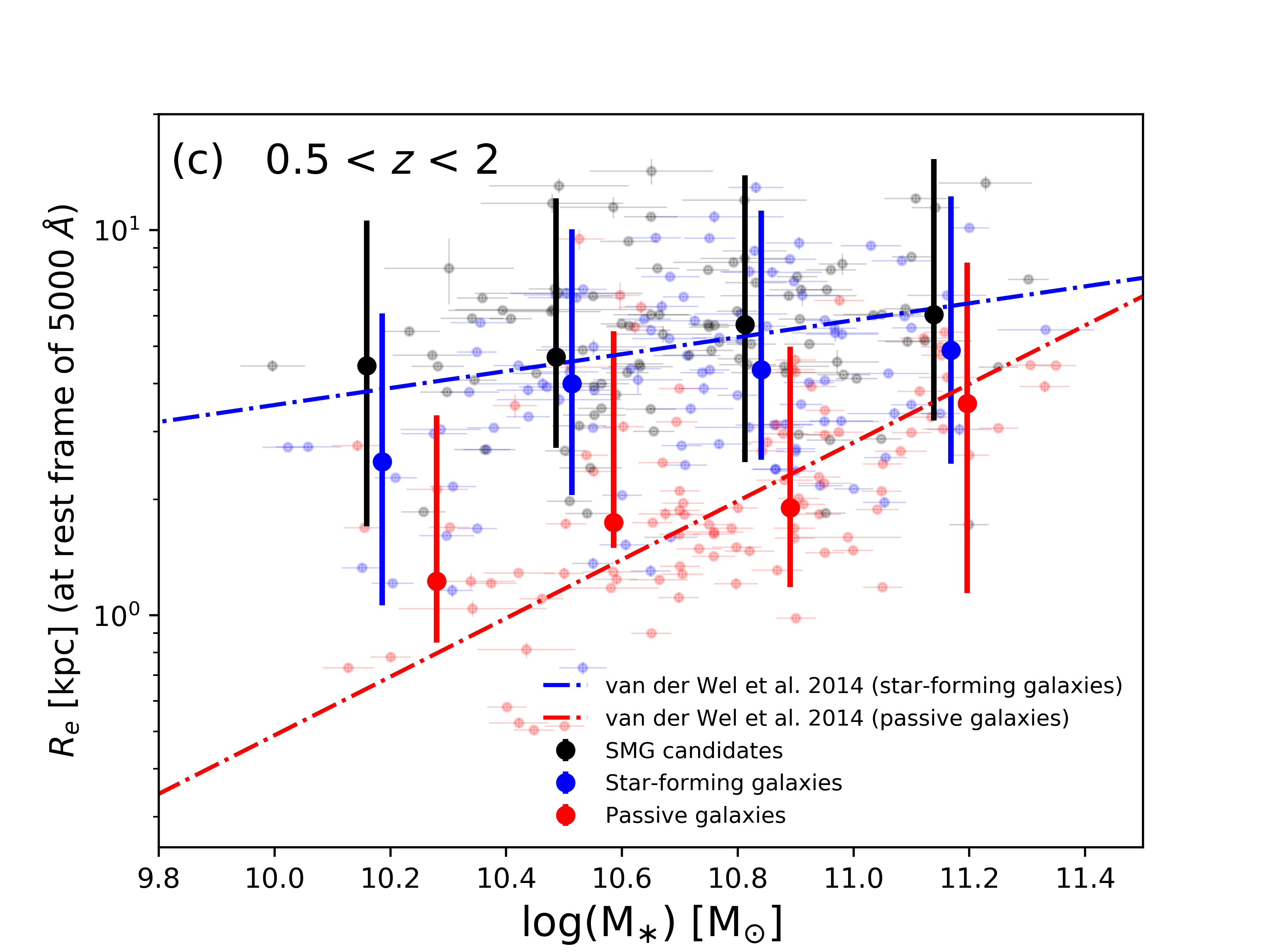}{0.5\textwidth}{}\hspace{-2em}
	\fig{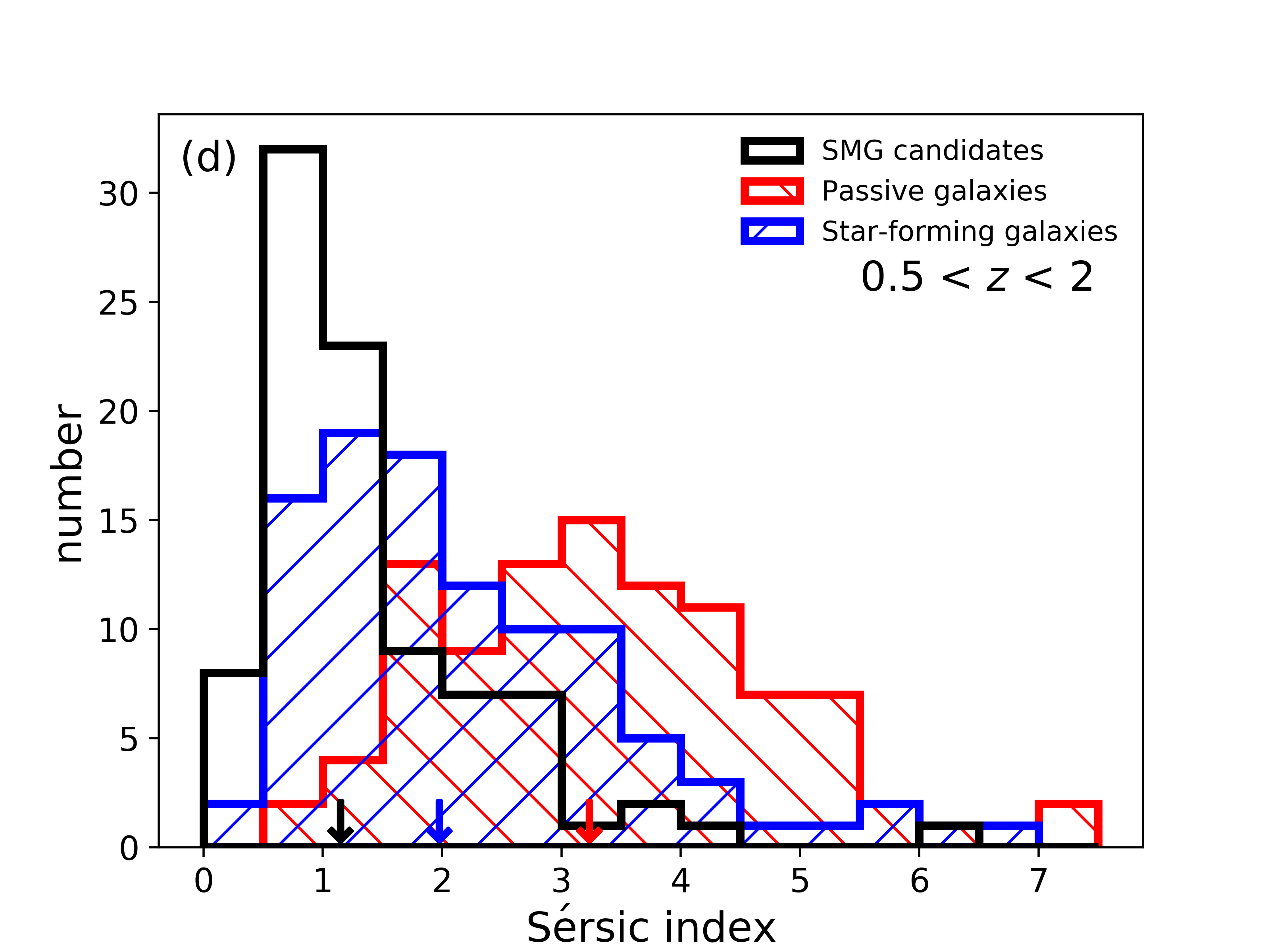}{0.5\textwidth}{}\hspace{-2em}}
	\caption{Physical properties of SMG candidates and comparison galaxy samples showing (a) normalized histograms of extinction, (b) normalized histograms of stellar population age, (c) the effective radius and stellar mass relations, and (d) histograms of S\'ersic index. In panels (a), (b), and (d), the median values are marked as downward arrows for the corresponding sample. In panel (c), the larger points show the running median of our sample and the $\pm1\sigma$ scatter, while the smaller circles show the effective radius and stellar mass relation of individual sources for the corresponding sample. In panel (c), we also show the best-fit size-mass relations of star-forming and passive galaxies at $z=1.25$ from \cite{vanderWel:2014aa}, which corresponds to the median redshifts of our sample. In summary, we find that the SMG candidates are dustier, younger, larger, and more disk-like than the comparison samples that are matched in redshift and stellar mass.}
	\label{fig:PhysicalProperties}
	\end{figure*}

\subsection{Physical properties of SMG candidates and comparison samples} \label{subsec:PhysicalProperties}

In this section, we compare some physical properties between our SMG candidates and comparison samples. The stellar mass, dust extinction, and age of the stellar population assembled in the model can be recovered from SED template fitting. In this work, we adopted the COSMOS2015 catalog values, estimated with the \texttt{LE PHARE} code. In short, they used a library of synthetic spectra generated using stellar population synthesis from \cite{Bruzual:2003aa} and assuming a \cite{Chabrier:2003aa} initial mass function. Both declining and delayed star-formation histories were used. The SEDs were generated for a grid of 42 ages in the range 0.05--13\,Gyr and the attenuation curve of \cite{Calzetti:2000aa} was applied to the templates with color excess in range of $E$($B$-$V$) = 0--0.7.

Our results show that the SMG candidates, which are expected to be dusty, tend to have higher values of $E$($B$-$V$), compared to star-forming and passive comparison samples (Figure \ref{fig:PhysicalProperties}(a)). A significant fraction of our SMG candidates have $E$($B$-$V$) $=0.7$ which is a cap on the maximum $E$($B$-$V$) value introduced in the \texttt{LE PHARE} SED fitting procedure. This implies that the extinction corrections, which are applied to the sources in the COSMOS2015 catalog, may insufficient and consequently their dust-corrected SFR estimations may be underestimated, particularly for the dusty population (see also \citealt{Casey:2013aa, Simpson:2014aa, Elbaz:2018aa, Dudzeviciute:2020aa}).

Figure \ref{fig:PhysicalProperties}(b) shows histograms of model stellar population age for our SMG candidates and comparison samples. It is worth noting that the age measurements derived from SED fitting alone should perhaps be treated with skepticism (see \citealt{Dudzeviciute:2020aa}), since the galaxy colors becomes redder with increasing extinction or age (the age-dust degeneracy; e.g. \citealt{Calzetti:2001aa, Pforr:2012aa}). This degeneracy might be more severe in the starburst systems such as SMGs \citep{Hainline:2011aa, Michaowski:2012aa, Simpson:2014aa}. On the other hand, the age measurements are highly dependent on the assumed star-formation history, in which the bursty star-formation models tend to produce the youngest ages, while the continuous star-formation models tend to produce the oldest ages \citep{Maraston:2010aa, Hainline:2011aa}. Nevertheless, the distribution of stellar population age in SMG candidates peaks at younger ages (Figure \ref{fig:PhysicalProperties}(b)), even though the overall distribution of stellar masses is similar to that in comparison star-forming and passive galaxies. This suggests that the SMG candidates maybe galaxies with more recent star formation (i.e., a higher proportion of young stars), since the amount of attenuation in the rest-frame ultraviolet or optical is so large that the SED fitting code will have to de-redden the optical SED to the youngest available stellar populations.

To investigate the spatial structure of our SMG candidates and comparison samples, we adopt the morphological properties of $\emph{Hubble Space Telescope}$ ($\emph{HST}$) $H_{\rm F160W}$-selected galaxies in the CANDELS /COSMOS field from \cite{vanderWel:2012aa}. We only use the sources with ``flag = 0" in this catalog, which consists of the objects with a good S\'ersic model fits \citep{Sersic:1963aa, Sersic:1968aa} measured by the \texttt{GALFIT} code \citep{Peng:2010aa}. According to \cite{vanderWel:2012aa}, the structural parameters can be measured with a precision and accuracy better than 10\% down to $H_{\rm F160W}$ $\simeq 24.5$\,mag$_{\rm AB}$. It is worth noting that our work here is similar to that in \cite{Chang:2018aa}. \cite{Chang:2018aa} used a sample of SCUBA-2 450-$\micron$-selected SMGs with $S_\mathrm{450\,\micron} \gt$ 2\,mJy, while we adopting machine-learned SMG candidates with $S_\mathrm{450\,\micron} \gt$ 4\,mJy (see \S\ref{sec:Justifications}). Our sample size is about 50\% larger than that in \cite{Chang:2018aa}, since we can use the entire catalog from \cite{vanderWel:2012aa} for our SMG candidates. In total, we have 91 SMG candidates, 102 star-forming galaxies, and 99 passive galaxies with robust S\'ersic model fit. 

Figure \ref{fig:PhysicalProperties}(c) shows the effective radius ($R_{\rm e}$) at a rest-frame wavelength of $\simeq5000\,\AA$ and stellar mass relations for our SMG candidates and the comparison samples. To determine the $R_{\rm e}$ value at a rest-frame wavelength of $\simeq5000\,\AA$, we follow the Equations 1 and 2 in \cite{vanderWel:2014aa}, which consider the wavelength dependence of $R_{\rm e}$ as a function of redshift and stellar mass. In Figure \ref{fig:PhysicalProperties}(c), we also show the best-fit size-mass relations of star-forming and passive galaxies at $z=1.25$ from \cite{vanderWel:2014aa}, which corresponds to the median redshifts of these sub-samples. The comparison passive galaxies are, on average, smaller than star-forming galaxies, which is consistent with earlier studies on both local and high-redshift samples \citep{Shen:2003aa, Ichikawa:2012aa, Newman:2012aa, FernandezLorenzo:2013aa, vanderWel:2014aa}. The median $R_{\rm e}$ of our SMG candidates ($5.5^{+0.3}_{-0.4}$\,kpc) is consistent with previous studies of ALMA follow-up at LABOCA 870-$\micron$-selected SMGs at $z \simeq 2$ ($R_{\rm e} = 4.4^{+1.1}_{-0.5}$\,kpc; \citealt{Chen:2015aa}), SCUBA-2 450-$\micron$-selected SMGs at $z=0.5$--1.5 ($R_{\rm e} = 4.9\pm0.3$\,kpc; \citealt{Chang:2018aa}), and ALMA follow-up at SCUBA-2 SMGs ($R_{\rm e} = 4.8\pm0.3$\,kpc; \citealt{Lang:2019aa}), within the errors, but somewhat higher than the measurement of radio-identified SMGs at $z \simeq 2$ ($R_{\rm e} = 2.8\pm0.4$\,kpc; \citealt{Swinbank:2010aa}).  \cite{Chen:2015aa} attributed this to the fact that \cite{Swinbank:2010aa} used shallower $\emph{HST}$-NICMOS images, which tend to show smaller sizes. Our result shows that the SMG candidates are significantly ($\simeq 3\sigma$) more extended than the comparison star-forming galaxies ($R_{\rm e} = 4.0^{+0.3}_{-0.2}$\,kpc). Even though we split our samples by stellar mass, we still find the trend that the SMG candidates are slightly larger than comparison star-forming galaxies in all mass bins (Figure \ref{fig:PhysicalProperties}(c)). The two-dimensional KS test (\texttt{python}-package \texttt{ks2d2s}; original estimation references from \citealt{Peacock:1983aa}) in the size-mass plane shows that the SMG candidates are significantly different from comparison star-forming galaxies ($p$-value = 0.006). We therefore confirm and strengthen the tentative findings in \cite{Chang:2018aa}, in which they showed that the 450- and 850-$\micron$-selected SMGs are marginally different from a stellar-mass- and SFR-matched star-forming sample.

The S\'ersic indices ($n_{\rm s}$) for our SMG candidates and comparison samples, which are from \cite{vanderWel:2012aa}, are shown as histograms in Figure \ref{fig:PhysicalProperties}(d). The median $n_{\rm s}$ of our SMG candidates ($1.1\pm0.1$) is consistent with previous studies, including $n_{\rm s}=1.4\pm0.8$ \citep{Swinbank:2010aa}, $n_{\rm s}=1.2\pm0.3$ \citep{Chen:2015aa}, $n_{\rm s}=1.1\pm0.1$ \citep{Chang:2018aa}, and $n_{\rm s}=1.0\pm0.2$ \citep{Lang:2019aa}. Our SMG candidates are statistically distinguishable, with a lower median of $n_{\rm s}$, compared to comparison star-forming ($n_{\rm s}=1.9\pm0.1$) and passive galaxies ($n_{\rm s}=3.2^{+0.2}_{-0.1}$). However, our result does not necessarily show that the SMG candidates are dominated by disk-like structures ($n_{\rm s} \simeq 1$), since most of the SMGs with low $n_{\rm s}$ ($\simeq 1$) are in fact visually classified as either irregulars or interacting systems \citep{Chen:2015aa}.

In summary, we find that the SMG candidates are dustier, younger, larger, and more disk-like than the comparison samples that are matched in redshift and stellar mass. With these differences in mind, in the next section we discuss their clustering properties.

\section{Clustering and Halo Mass} \label{sec:Clustering}
We now investigate the standard two-point clustering statistics that can quantitatively measure the large-scale structure of the Universe and trace the amplitude of galaxy clustering as a function of scale. The clustering measurements can further allow us to infer the dark-matter halo mass and to constrain the evolution of clustering. 

\begin{deluxetable*}{cccccc}
\tablecaption{\label{tbl:Clustering} Results of Clustering Analyses}
\tablehead{  \colhead{Sample} & \colhead{Redshift} & \colhead{$N_{\rm s}$\tablenotemark{a}} &  \colhead{$b$} &  \colhead{$r_{\rm 0}$} & \colhead{log($M_{\rm halo}$)}  \\
 & & & & \colhead{($h^{-1}$\,Mpc)} & \colhead{($h^{-1}\,\rm M_{\sun}$)} }
\startdata
SMG candidates & $0.5 < z < 1.0 $ & 1112 & 2.4$^{+0.3}_{-0.3}$ & 8.1$^{+1.0}_{-1.1}$ & 13.4$^{+0.2}_{-0.2}$ \\
			  			  & $1.0 < z < 2.0 $ & 2205 & 3.9$^{+0.4}_{-0.4}$ & 9.8$^{+1.1}_{-1.1}$ & 13.4$^{+0.1}_{-0.2}$ \\ 
			  			  & $2.0 < z < 3.0 $ & 1518 & 5.9$^{+0.7}_{-0.9}$ & 10.9$^{+1.5}_{-1.7}$ & 13.2$^{+0.2}_{-0.2}$ \\\hline
Passive galaxies 			  & $0.5 < z < 1.0$ & 1112 & 3.3$^{+0.2}_{-0.2}$ & 11.5$^{+0.8}_{-0.8}$ & 13.9$^{+0.1}_{-0.1}$ \\
			  			 & $1.0 < z < 2.0$ & 1909 & 4.3$^{+0.4}_{-0.5}$ & 10.8$^{+1.2}_{-1.3}$ & 13.5$^{+0.1}_{-0.2}$\\\hline
Star-forming galaxies 		 & $0.5 < z < 1.0$ & 1090 & 2.5$^{+0.3}_{-0.3}$ & 8.4$^{+1.0}_{-1.0}$ & 13.5$^{+0.1}_{-0.2}$ \\
			 	  		 & $1.0 < z < 2.0$ & 1993 & 4.1$^{+0.4}_{-0.5}$ & 10.4$^{+1.2}_{-1.3}$ & 13.5$^{+0.1}_{-0.2}$ \\
\enddata
\tablenotetext{a} { Sample sizes of our samples in the corresponding redshift bins. }
\end{deluxetable*}

	\begin{figure*}
	\centering
	\includegraphics[width=0.9\paperwidth]{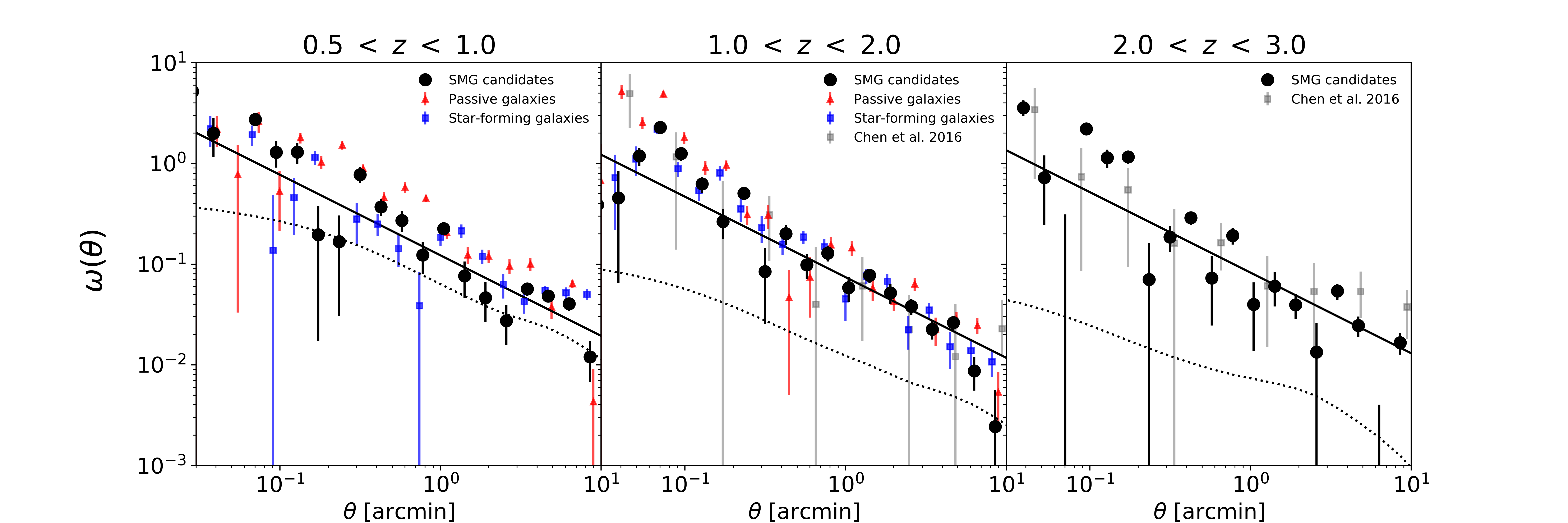}
	\caption{Two-point auto-correlation function of our SMG candidates and the comparison samples at $0.5<z<3.0$. The data points are offset horizontally, to avoid confusion. The dotted curves show the auto-correlation functions of the dark matter in the corresponding redshift bins. }
	\label{fig:AutoCorrelation}
	\end{figure*}

	\begin{figure*}
	\centering
	\includegraphics[width=0.9\paperwidth]{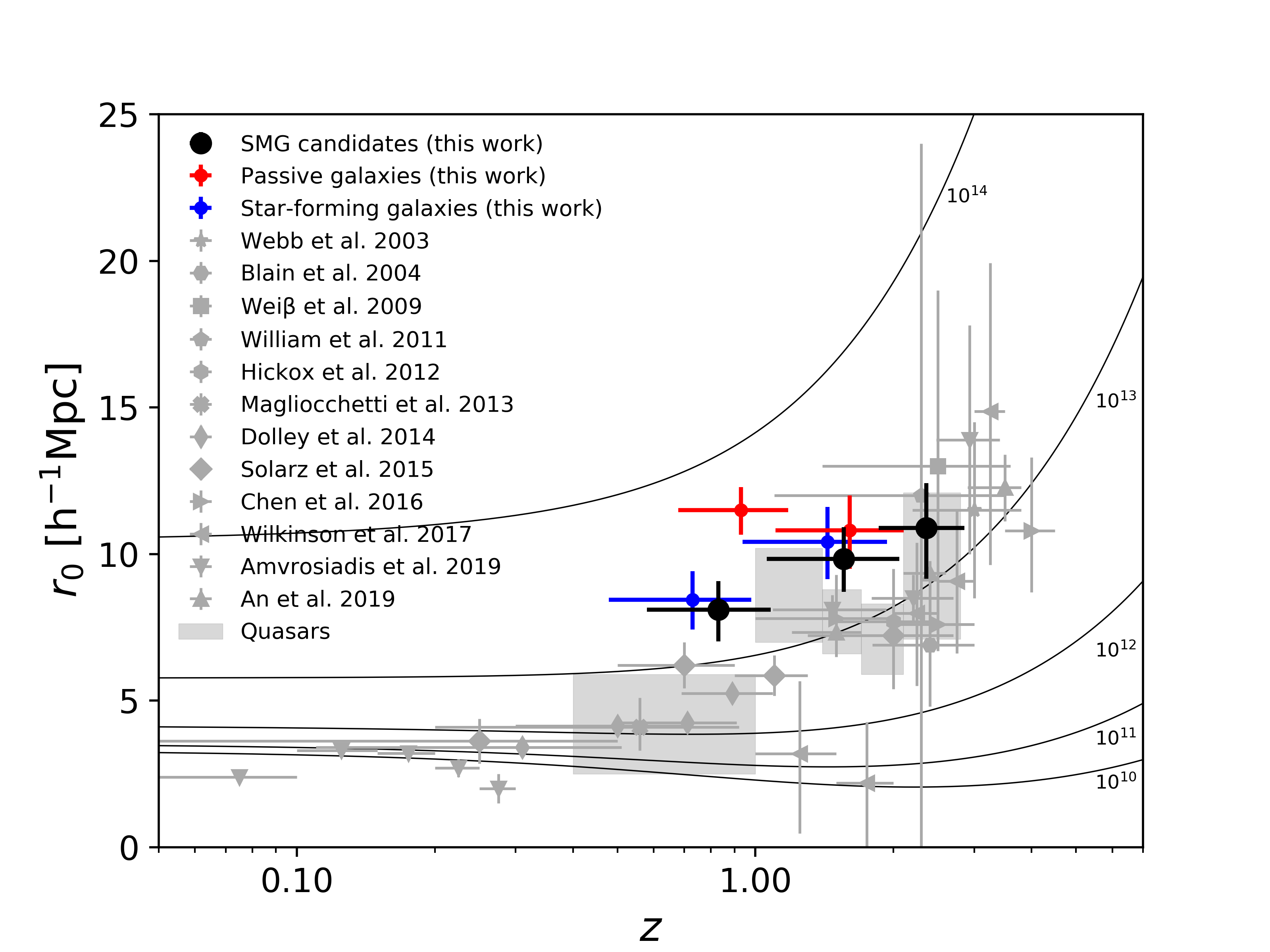}
	\caption{Redshift evolution of the clustering length $r_{\rm 0}$ for our SMG candidates and the comparison samples. The data points are slightly offset horizontally, to avoid overlap. We find no evidence that the clustering signal of SMG candidates exhibits an evolution with redshift. SMG candidates reside in a typical halo mass of $\simeq (2.0\pm0.5) \times10^{13}\,h^{-1}\,\rm M_{\sun}$ across the redshift range of $0.5<z<3$.  We also show the estimated $r_{\rm 0}$ of 24-$\micron$-selected galaxies \citep{Dolley:2014aa, Solarz:2015aa}, SMGs \citep{Webb:2003ab, Blain:2004aa, WeiB:2009aa, Williams:2011aa, Hickox:2012aa, Magliocchetti:2013aa, Chen:2016ab, Wilkinson:2017aa, Amvrosiadis:2019aa, An:2019aa}, and quasars \citep{Myers:2006aa, Porciani:2006aa, Shen:2007aa, Eftekharzadeh:2015aa} in literature for comparison.}
	\label{fig:ClusteringLength}
	\end{figure*}

\subsection{Two point auto-correlation function} \label{subsec:TwoPointACF}
We measure the two-point galaxy auto-correlation function by using the \cite{Landy:1993aa} estimator:

\begin{equation}
\omega{(\theta)}  = \frac{ DD(\theta) - 2DR(\theta) + RR(\theta) } { RR(\theta) },
\label{eq:LandySzalay}
\end{equation}
where $DD(\theta)$, $DR(\theta)$, and $RR(\theta)$ are the normalized number of galaxy-galaxy, galaxy-random, and random–random pairs, respectively, counted within a given angular separation bin of $\theta\pm\delta\theta/2$. The $DR(\theta)$ and $RR(\theta)$ are normalized to the same total number of pairs as $DD(\theta)$, with $DR(\theta) = \frac{ (N_{\rm D} - 1) } { 2N_{\rm R} } N_{\rm DR}(\theta)$, and $RR(\theta) = \frac{ N_{\rm D}(N_{\rm D} - 1) } { N_{\rm R} (N_{\rm R} - 1) } N_{\rm RR}(\theta)$, where the $N_{\rm D}$ and $N_{\rm R}$ are the number of sources in the galaxy and random samples, respectively; while the $N_{\rm DR}(\theta)$ and $N_{\rm RR}(\theta)$ are the original counts of galaxy-random and random–random pairs, respectively. We adopted ten times as many random points as the number of our galaxy.

The projected angular two-point auto-correlation function, $\omega{(\theta)}$, can generally be described as a power law, 
\begin{equation}
\omega{(\theta)_{\rm mod}}  = A\theta^{-\delta}, 
\label{eq:Omega}
\end{equation}
where $A$ is the clustering amplitude and $\delta$ is the slope of the correlation function. The value $\delta=0.8$ has been found to be appropriate for both observations and theories at the physical separation of $\simeq 0.1$--$10\,h^{-1}$\,Mpc (e.g., \citealt{Peebles:1980aa, Davis:1983aa, Zehavi:2005aa, Coil:2007aa, Coil:2008aa}). This statement still holds for the galaxy samples at redshift up to $\simeq 5$ ($\delta=0.8$--1.1; \citealt{Ouchi:2005aa, Coil:2006aa, Kashikawa:2006aa, Lee:2006aa, Hildebrandt:2009aa, Durkalec:2015aa, Harikane:2016aa, Jose:2017aa}).

Measurement of $\omega{(\theta)}$ could be biased, since our sample is located in a single, area-limited region that may not be representive of the true underlying mean density over the whole sky. The observed $\omega{(\theta)}$ is usually biased low than the true value $\omega{(\theta)_{\rm mod}}$, 

\begin{equation}
\omega{(\theta)}  = \omega{(\theta)}_{\rm mod} - \rm IC,
\label{eq:ObsOmega}
\end{equation}
by an additive factor of IC known as the integral constraint. In practice, IC can be numerically estimated (e.g., \citealt{Infante:1994aa, Roche:1999aa, Adelberger:2005aa}) over the survey geometry using the random-random pairs under the form 

\begin{equation}
{\rm IC}  = \frac{ \sum_{i} N_{\rm RR}(\theta_{i}) \omega{(\theta_{i})}_{\rm true} } { \sum_{i} N_{\rm RR}(\theta_{i}) }.
\label{eq:IC}
\end{equation}

We repeated the estimate of Equation~\ref{eq:LandySzalay} 25 times. We calculated the variance, mean $\omega(\theta)$, and the mean $N_{\rm rr}$ using these 25 estimates in Equation~\ref{eq:IC}. We then employed a $\chi^2$ minimization in Equation~\ref{eq:ObsOmega} to derive the best fit $\omega(\theta)_{\rm true}$ over scales of 0$\farcm$1--6$\arcmin$ ($\simeq 0.1$--6\,$h^{-1}$\,Mpc at $z$ = 2 for comoving distance). The IC-corrected $\omega(\theta)$ is shown as black symbols in Figure~\ref{fig:AutoCorrelation}. 

The errors in the clustering amplitude of $\omega(\theta)_{\rm true}$ are expected to be underestimated, since the variance only accounts for the shot noise from the sample of the random points and the Poisson uncertainties of the DD counts, and does not include the field-to-field variance (often called cosmic variance) of a field-limited survey. To quantify the uncertainty more realistically, we employed the ``delete one jackknife" resampling method (see also \citealt{Scranton:2002aa, Zehavi:2002aa, Zehavi:2005aa, Norberg:2009aa}) to estimate the covariance matrix for the auto-correlation function measurements. The principle of the jackknife method is to first divide the dataset into $N$ independent subsamples and then one copy of the subsamples is omitted systematically at a time similar to the $k$-fold cross-validation used in the machine learning. Therefore, the resampling dataset consists of $N_{\rm sub} - 1$ remaining sub-area, with area $(N_{\rm sub} - 1)/N_{\rm sub}$ times the full area of the original dataset. The covariance matrix of $N$ independent realizations can be obtained as
\begin{equation*}
C_{ij} =  \frac{N_{\rm sub} - 1} {N_{\rm sub}} \sum^{N_{\rm sub}}_{k=1} (\omega(\theta_i)^k - \overline{\omega(\theta_i)}) (\omega(\theta_j)^k - \overline{\omega(\theta_j)}),
\end{equation*}
where $\omega(\theta_{i,j})^k$ is the auto-correlation function measured with the $k^{\rm th}$ area removed and $\overline{\omega(\theta_i)}$ is the average auto-correlation function of the jackknife realizations. In practice, we divide our sample into $N_{\rm sub}=13$ nearly equal-size stripe-shaped sub-areas. We verify that the shapes of the sub-area do not affect the jackknife results significantly. To determine the best fit power-law model (Equation \ref{eq:Omega}) for each correlation function, we perform a $\chi^{2}$ minimization, where 
\begin{equation*}
\chi^2 =  \sum_{i} \sum_{j} (\omega(\theta_i)^k - \omega(\theta_i)_{\rm mod}) C_{ij}^{-1} (\omega(\theta_j)^k - \omega(\theta_j)_{\rm mod}).
\end{equation*}
The $1\sigma$ error is estimated based on finding where $\Delta\chi^2=1$ \citep{Avni:1976aa}.

\subsection{Dark-matter halo mass} \label{subsec:HaloMass}

To quantify the underlying dark matter halo mass ($M_{\rm halo}$) of our sample, we first need to compute the galaxy bias ($b$), which can be defined as the square root of the ratio of the two-point correlation function of the galaxies relative to the dark matter:
\begin{equation*}
b = \left(\frac{\omega(\theta)}{\omega(\theta)_{\rm DM}}\right)^{\frac{1}{2}},
\end{equation*}
where the $\omega(\theta)_{\rm DM}$ is the projected angular two-point correlation function of the dark matter. To reproduce the clustering model in dark matter, we use the \texttt{HALOFIT} code from \cite{Smith:2003aa} with improved fitting formulae provided by \cite{Takahashi:2012aa}, which can predict the nonlinear and dimensionless power spectrum of dark matter $\Delta^2(k) =  k^3P(k)/(2\pi)^2$ for a wide range of cold dark matter (CDM) cosmologies. The Fourier transform of the two-point correlation function is the power spectrum. We then project the power spectrum into the angular correlation function by using Limber's equation \citep{Limber:1953aa, Peebles:1980aa, Peacock:1991aa, Baugh:1993aa}, specifically via the Equation A6 in \cite{Myers:2007aa}. The $\omega(\theta)_{\rm DM}$ profiles at $z=0.5$--1, $z=1$--2, and $z=2$--3 are shown as dotted curves in Figure \ref{fig:AutoCorrelation}. We fit a single $b$ parameter from the observed galaxy correlation function and the dark matter correlation function by minimizing the $\chi^{2}$ on the scales of 0$\farcm$1--6$\arcmin$. Finally, we convert the $b$ to $M_{\rm halo}$ using the ellipsoidal collapse model of \citet{Sheth:2001aa}.

In the case of the small-angle approximation ($\theta \ll 1$\,rad) and assuming no clustering evolution over the redshift bin, we can de-project the angular auto-correlation function to the power spectrum by inverting Limber's equation (e.g., \citealt{Myers:2006aa, Hickox:2012aa}; see \citealt{Peebles:1980aa} for full detivation) and further estimate the clustering scale length ($r_{\rm 0}$) as the following: 
\begin{equation}
A = H_\gamma \frac{\int^{\infty}_{0} (dN/dz)^2 E_{z} \chi^{1-\gamma}dz} {[(dN/dz)dz]^{2}} r^{\gamma}_{0},
\label{eq:A}
\end{equation}
where $H_\gamma = \Gamma(1/2)\Gamma([\gamma-1]/2)/\Gamma(\gamma/2)$, $\Gamma$ being the gamma function, $\gamma=\delta+1$ (=1.8 in this work), $\chi$ is the radial comoving distance, $dN/dz$ is the redshift selection function, and $E_{z}=H_{z}/H_{0}=dz/d\chi$ ($H_{z}^{2} = H_{0}^{2}[\Omega_{\rm m}(1+z)^{3}+\Omega_{\rm \Lambda}]$). 

According to the formalism of \citet{Peebles:1980aa}, $r_{\rm 0}$ is related to $b$ via
\begin{equation}
r_{\rm 0} = 8 \left(\frac{\Delta^{2}_{8}}{C_{\gamma}}\right)^{1/\gamma} = 8 \left( \frac{b^{2}\sigma^{2}_{8}D^{2}}{C_{\gamma}}\right)^{1/\gamma},
\label{eq:r0}
\end{equation}
where $C_{\gamma}=72/(3-\gamma)(4-\gamma)(6-\gamma)2^{\gamma}$, $\sigma_{8}$ is the amplitude of matter clustering (= 0.83; \citealt{Planck:2014aa}), and $\Delta_{8}$ is the clustering strength of dark matter haloes more massive than stellar mass $M$ at redshift $z$, which can be defined as $\Delta_{8}=b(M,z)\sigma_{8}D(z)$. The function $D(z)$ is the growth factor of linear fluctuations in the dark-matter distribution, which can be computed from 
\begin{equation*}
D(z) = \frac{5\Omega_{\rm m}E_{z}}{2}\int^{\infty}_{z} \frac{1+y}{E_{z}^{3}(y)}dy.
\end{equation*}
In our subsequent analysis, the redshift value is assumed to be the median of the distribution of the sources. The results are summarized in Table \ref{tbl:Clustering}. 

\subsection{Clustering signals} \label{subsec:ClusteringSignals}

Figure \ref{fig:ClusteringLength} shows the values of $r_{\rm 0}$ as a function of redshift for our SMG candidates and for comparison samples. We also plot the measurements from the literature for 24-$\micron$-selected galaxies \citep{Dolley:2014aa, Solarz:2015aa}, 100-$\micron$-selected $\emph{Herschel}$ SMGs \citep{Magliocchetti:2013aa}, 250-$\micron$-selected $\emph{Herschel}$ sources \citep{Amvrosiadis:2019aa}, 850-$\micron$-selected SMGs \citep{Webb:2003ab, Blain:2004aa, WeiB:2009aa, Williams:2011aa, Hickox:2012aa, Chen:2016ab, Wilkinson:2017aa, An:2019aa}, and quasars \citep{Myers:2006aa, Porciani:2006aa, Shen:2007aa, Eftekharzadeh:2015aa}. The measured $r_{\rm 0}$ (or $b$) of our SMG candidates and the comparison samples decline with decreasing redshift (Table \ref{tbl:Clustering}). This trend is expected, since dark matter clustering evolves rapidly as the Universe evolves with time (dotted curves in Figure \ref{fig:AutoCorrelation}), resulting in the measured value of $r_{\rm 0}$ (or $b$) of a biased population will decrease. The preceding analysis only illustrates the clustering signals of our sample, so combining the knowledge of clustering signals and halo masses of our samples is more meaningful. Our SMG candidates reside in a halo with a typical mass of $\simeq (2.0\pm0.5) \times10^{13}\,h^{-1}\,\rm M_{\sun}$ across the redshift range $0.5<z<3$. In general, passive galaxies have stronger clustering signals than the comparison star-forming galaxies and SMG candidates, indicating that passive galaxies preferentially reside in more massive halos compared to star-forming galaxies and SMG candidates at fixed stellar mass and redshift (see also \citealt{Hartley:2008aa, Hartley:2010aa, McCracken:2010aa, Lin:2012aa, Lin:2016aa, Sato:2014aa, Ji:2018aa}). On the other hand, there is no significant difference between the SMG candidates and the comparison star-forming galaxies for the same stellar mass cut, suggesting that these two populations reside in similar mass halos. Similar trends can also be found in earlier studies of galaxy samples at $1.5<z<2.5$ \citep{Bethermin:2014aa} and at $1<z<5$ \citep{Chen:2016ab, An:2019aa}. This result implies that merging events may not be the only triggering mechanism for SMGs, since we expect that more biased regions will result in higher merging and/or interaction rates \citep{Lin:2010aa, deRavel:2011aa, Sobral:2011aa}. However, further studies are still needed to address the question of what mechanisms increase the dust obscuration within galaxies. 

At $z=1$--3, the clustering measured for our SMG candidates are in broad agreement with the previous studies, except that \cite{Wilkinson:2017aa} found weaker clustering strength at $z \simeq 1$--2, although with large uncertainties (so not significantly different from our results). We find no strong evidence that the halo masses of SMG candidates exhibit any evolution with redshift. This is in agreement with what was found in \cite{Chen:2016ab}, \cite{Amvrosiadis:2019aa}, and \cite{An:2019aa}, but contrary to what was suggested in \cite{Wilkinson:2017aa}, in which they found that SMG activity seems to be shifting to less massive halos; consistent with an early downsizing scenario. We attribute this to their large measurement uncertainties or the different methodologies that are adopted in the clustering analyses. We note, however, that the term downsizing is a relative notion. While we do not observe a significant decrease of halo masses of 450-$\micron$ SMGs with decreasing redshifts, which is why we say we do not observe downsizing, we are ultimately comparing halo masses at different redshifts. Given the same mass, halos at higher redshifts are expected to grow into more massive halos at the present day. From this perspective, our results support the downsizing scenario such that the 450-$\micron$ SMGs at lower redshifts are formed within halos that are on average smaller in mass at the present day.

Our result provides a meaningful constraint on the clustering amplitude of SMGs at $z\simeq0.5$--1, a key redshift range where downsizing effects are expected to take place. At $z\simeq0.5$--1, the clustering signal of our SMG candidates appears to be a little higher than in the previous studies of \cite{Magliocchetti:2013aa, Dolley:2014aa, Solarz:2015aa}, although there are large uncertainties. It is worth noting that the majority of the sources in the aforementioned studies represent a fainter population with $L_{\rm IR} \simeq 10^{11}\,\rm L_{\sun}$. On the other hand, our SMG candidates are bright at 450\,$\micron$ ($\simeq 5$\,mJy; see \S\ref{sec:Justifications}) which corresponds to $L_{\rm IR} \simeq 10^{12}\,\rm L_{\sun}$ at $z=0.5$--1 if we convert the 450-$\micron$ flux density into the total $L_{\rm IR}$ by using the average ALESS 870-$\micron$ SEDs \citep{da-Cunha:2015aa}. Therefore, a stronger clustering signal is expected in our sample, since the galaxies with higher $L_{\rm IR}$ tend to have stronger clustering \citep{ Dolley:2014aa, Toba:2017aa}.

By comparing the $z<1$ measurements and those at $z>1$, along with the quasars, there is a clear trend for clustering length to increase with redshift at earlier epochs, which is consistent with downsizing behavior. The question now arises: Are the SMGs at $z<1$ the same dusty population as those at $z>1$? The dusty star-forming galaxy population seems to have lower $L_{\rm IR}$ at low redshift (e.g., \citealt{Magliocchetti:2013aa, Dolley:2014aa, Solarz:2015aa}), which prevents us from making a direct comparison with the SMGs at high redshift. Nevertheless, our understanding of the clustering properties in SMGs is still far from complete. The study of the redshift evolution of the clustering properties in SMGs with comparable $L_{\rm IR}$ at a similar wavelength is required to solve this problem. We expect this situation to be improved with the next generation sub-millimeter telescope such as Atacama Large Aperture Submm/mm Telescope (AtLAST; \citealt{Klaassen:2019aa}). A future AtLAST survey with increased sensitivity at 350\,$\micron$ and larger field-of-view ($\simeq 1\degr$) will detect the fainter population ($L_{\rm IR}\simeq10^{11}\,\rm L_{\sun}$) at larger scales.

\section{Summary} \label{sec:Summary}

By combining SCUBA-2 data from the ongoing JCMT Large Program STUDIES and the archive in the CANDELS/COSMOS field, we have obtained an extremely deep 450-$\micron$ image (1$\sigma$ = 0.56\,mJy\,beam$^{-1}$) covering $\simeq 300$\,arcmin$^{2}$: by far the deepest image ever observed at 450 $\micron$. We obtain a sample of 221 450-$\micron$-selected SMGs from this image, however, the sample size is too small to meaningfully study the redshift evolution of the clustering of the population. 

We select a robust (S/N $\geqslant 4$) and flux-limited ($\geqslant 4$\,mJy) sample of 164 450-$\micron$-selected SMGs that have $K$-band counterparts in the COSMOS2015 catalog identified based on radio or mid-infrared imaging, which allows us to employ their optical and near-infrared colors. Ultilizing this SMG sample and the 4705 $K$-band-selected non-SMGs lying within the $\leqslant 1$\,mJy\,beam$^{-1}$ noise level region of the 450-$\micron$ image as a training set, we develop a machine-learning classifier to identify SMG candidates in the full COSMOS field. We employ the $K$-band magnitudes and color-color pairs based on the thirteen-broad-band photometry measurements ($uBVri^{+}z^{++}JHK$[3.6][4.5][5.8][8.0]) available in this field for the machine-learning algorithm. Our main findings are the following.

\begin{itemize}

\item Our trained classifier labels 6182 SMG candidates in the wider COSMOS field from the COSMOS2015 catalog with $m_{\rm K}<24.5$\,mag$_{\rm AB}$ across an effective area of 1.6\,deg$^{2}$. 

\item The number density, VLA 3\,GHz and/or MIPS 24\,$\micron$ detection rates, redshift and stellar-mass distributions, and the stacked 450-$\micron$ flux densities of the SMG candidates across the COSMOS field agree with the measurements made in the much smaller CANDELS field, all supporting the effectiveness of the classifier. The high completeness ($76\%\pm7\%$) and precision ($82\%\pm7\%$) of our SMG candidates as judged from their detection in longer wavelength ALMA observations further supports our machine-learning algorithm.

\item We found that the SMG candidates tend to have higher reddening compared to comparison star-forming and passive galaxies that are matched in redshift and stellar mass. The SMG candidates also have younger stellar population ages than the comparison star-forming and passive galaxies, even though their overall distribution of stellar masses is similar. This suggests that the SMG candidates may have more recent star formation and consequently have a higher proportion of young stars. 

\item The SMG candidates have a median effective radius of $5.5^{+0.3}_{-0.4}$\,kpc and a median S\'ersic index of $1.1\pm0.1$. These measurements are consistent with previous studies within the uncertainties. Our results show that SMG candidates are significantly more extended and more disk-like than the comparison star-forming and passive galaxies. 

\item We measured the two-point autocorrelation function of the SMG candidates from $z=3$ down to $z=0.5$, and found that they reside in halos with masses of $\simeq (2.0\pm0.5) \times10^{13}\,h^{-1}\,\rm M_{\sun}$ across this redshift range. However, we do not find evidence of downsizing that has been suggested by other recent observational studies. 

\end{itemize}

\acknowledgments

We thank the JCMT/EAO staff for observational support and the data/survey management and the anonymous referee for comments that significantly improved the manuscript.
C.C.C. acknowledges support from the Ministry of Science and Technology of Taiwan (MOST 109-2112-M-001-016-MY3).
C.F.L. and C.C.C. acknowledge grant support from SSDF-(ST)18/27/E.
C.F.L., W.H.W., and Y.Y.C. acknowledge grant support from the Ministry of Science and Technology of Taiwan (105-2112-M-001-029-MY3 and 108-2112-M-001-014-).
I.R.S. acknowledges support from STFC (ST/P000541/1). 
L.C.H. acknowledges support from the National Science Foundation of China (11721303 and 11991052) and National Key R\&D Program of China (2016YFA0400702).
M.J.M. acknowledges the support of the National Science Centre, Poland through the SONATA BIS grant 2018/30/E/ST9/00208. 
M.P.K. acknowledges support from the First TEAM grant of the Foundation for Polish Science No. POIR.04.04.00- 00-5D21/18-00.
S.E.H. is supported by Basic Science Research Program through the National Research Foundation of Korea funded by the Ministry of Education (2018R1A6A1A06024977).
Y.A. acknowledges financial support by NSFC grant 11933011.
Y.G.'s research is supported by National Key Basic Research and Development Program of China (grant No. 2017YFA0402704), National Natural Science Foundation of China (grant Nos. 11861131007, 11420101002), and Chinese Academy of Sciences Key Research Program of Frontier Sciences (grant No. QYZDJSSW-SLH008).

The James Clerk Maxwell Telescope is operated by the East Asian Observatory on behalf of the National Astronomical Observatory of Japan; the Academia Sinica Institute of Astronomy and Astrophysics; the Korea Astronomy and Space Science Institute; and the Operation, Maintenance and Upgrading Fund for Astronomical Telescopes and Facility Instruments, budgeted from the Ministry of Finance (MOF) of China and administrated by the Chinese Academy of Sciences (CAS), as well as the National Key R\&D Program of China (No. 2017YFA0402700). Additional funding support is provided by the Science and Technology Facilities Council of the United Kingdom and participating universities in the United Kingdom and Canada.
The submillimeter observations used in this work include the S2COSMOS program (program code M16AL002), STUDIES program (program code M16AL006), S2CLS program (program code MJLSC01) and the PI program of \citet[][program codes M11BH11A, M12AH11A, and M12BH21A]{Casey:2013aa}.

\software{ GALFIT \citep{Peng:2010aa}, LE PHARE \citep{Arnouts:1999aa, Ilbert:2006aa}, scikit-learn \citep{Pedregosa:2011aa}, Scipy \citep{Jones:2001aa}, PICARD \citep{Jenness:2008aa}, SExtractor \citep{Bertin:1996aa}, SMURF \citep{Chapin:2013aa}}

\appendix
\section{Clustering results of SMG candidates identified by other algorithms}\label{sec:ClusteringResultsOther}

We adopt the 450-$\micron$ SMG candidates that are identified by decision tree (better recall) and random forest (better precision), and estimate their clustering signals by using the same procedures in \S\ref{sec:Clustering}. The results are shown in Table \ref{tbl:ClusteringOther}. The decision tree and random forest label 14440 and 1866 450-$\micron$ SMG candidates, respectively, from the COSMOS2015 catalog at redshift $0.5<z<3.0$. The larger number of 450-$\micron$ SMG candidates found using the decision tree algorithm is expected, since the higher recall will select more sources, but as a tradeoff, the precision decreases. In contrast, the situation is reserved for random forest algorithm. The stacked fluxes of the 450-$\micron$ SMG candidates identified by the decision tree and random forest methods are ($3.0\pm0.1$)\,mJy and ($6.4\pm0.3$)\,mJy, respectively. As expected we find a lower stacked flux with the decision tree and a higher one with the random forest, essentially reflecting their precision. It is worth noting that the random forest algorithm only labels $\sim100$ 450-$\micron$ SMG candidates in the redshift bin of $0.5<z<1.0$, and consequently we do not have sufficient data points for the clustering analyses in this redshift bin. Nevertheless, the clustering signals do not show significantly differences from the results of XGBoost identified SMG candidates (Table \ref{tbl:Clustering}), indicating that our final results do not strongly depend upon the method we choose.

\begin{deluxetable*}{cccccc}
\tablecaption{\label{tbl:ClusteringOther} Clustering Results of 450-$\micron$ SMG Candidates Identified by Decision Tree and Random Forest Algorithms.}
\tablehead{  \colhead{Sample} & \colhead{Redshift} & \colhead{$N_{\rm s}$\tablenotemark{a}} &  \colhead{$b$} &  \colhead{$r_{\rm 0}$} & \colhead{log($M_{\rm halo}$)}  \\
 & & & & \colhead{($h^{-1}$\,Mpc)} & \colhead{($h^{-1}\,\rm M_{\sun}$)} }
\startdata
450-$\micron$ SMG candidates (Decision Tree) & $0.5 < z < 1.0 $ & 2406 & 2.4$^{+0.2}_{-0.2}$ & 8.1$^{+0.7}_{-0.8}$ & 13.4$^{+0.1}_{-0.1}$ \\
						  & $1.0 < z < 2.0 $ & 8919 & 3.2$^{+0.2}_{-0.2}$ & 8.0$^{+0.5}_{-0.5}$ & 13.1$^{+0.1}_{-0.1}$ \\
   						  & $2.0 < z < 3.0 $ & 3115 & 6.2$^{+0.5}_{-0.5}$ & 11.5$^{+1.0}_{-1.0}$ & 13.3$^{+0.1}_{-0.1}$ \\\hline
450-$\micron$ SMG candidates (Random Forest) & $0.5 < z < 1.0 $ & 143 & ...  &  ... & ... \\
						    & $1.0 < z < 2.0 $ & 730 & 5.3$^{+1.0}_{-1.3}$ & 13.8$^{+3.0}_{-3.6}$ & 13.8$^{+0.2}_{-0.4}$ \\
						    & $2.0 < z < 3.0 $ & 993 & 6.0$^{+1.3}_{-1.7}$ & 11.0$^{+2.8}_{-3.5}$ & 13.2$^{+0.3}_{-0.5}$ \\
\enddata
\tablenotetext{a} { Sample sizes of our samples in the corresponding redshift bins. }
\end{deluxetable*}

\bibliography{references.bib}

\end{document}